\documentclass{article}

\usepackage{arxiv}

\usepackage[utf8]{inputenc} 
\usepackage[T1]{fontenc}    
\usepackage{hyperref}       
\usepackage{url}            
\usepackage{booktabs}       
\usepackage{amsfonts}       
\usepackage{amsmath}
\usepackage{nicefrac}       
\usepackage{microtype}      
\usepackage{lipsum}		
\usepackage{graphicx}
\usepackage{varwidth}
\usepackage[numbers,sort&compress]{natbib}
\usepackage{doi}

\usepackage{tikz}
\usetikzlibrary{positioning}
\usetikzlibrary{calc}
\usepackage{subcaption}
\usepackage{program}
\usepackage{algpseudocode}
\usepackage{algorithm}
\usepackage{algorithmicx}
\usepackage{siunitx}
\usepackage{mathtools}
\usepackage[np, autolanguage]{numprint}

\definecolor{myblue}{HTML}{52afe6}
\definecolor{mygreen}{HTML}{239b56}

\title{Seamless Integration of Analysis and Design: Automatic CAD Reconstruction of Post-Analysis Geometries}

\author{ \href{https://orcid.org/ 0000-0001-7795-9760 }{\includegraphics[scale=0.06]{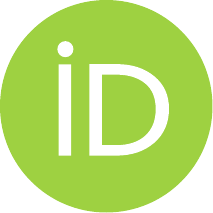}\hspace{1mm}Sebastian Hube}\\
	Chair for Computational Analysis of Technical Systems (CATS)\\
	RWTH Aachen University\\
	Schinkelstr. 2, 52062 Aachen, Germany \\
	\texttt{hube@cats.rwth-aachen.de} \\
	\And
	\href{https://orcid.org/ 0000-0002-4050-1841 }{\includegraphics[scale=0.06]{orcid.pdf}\hspace{1mm}Roxana Pohlmann  } \\
	Institute of Lightweight Design and Structural Biomechanics (E317)\\
	TU Wien\\
	Gumpendorfer Str. 7, 1060 Vienna, Austria \\
	\texttt{roxana.pohlmann@tuwien.ac.at} \\
	\And
	\href{https://orcid.org/0000-0002-4474-1666}{\includegraphics[scale=0.06]{orcid.pdf}\hspace{1mm}Stefanie Elgeti} \\
	Institute of Lightweight Design and Structural Biomechanics (E317)\\
	TU Wien\\
	Gumpendorfer Str. 7, 1060 Vienna, Austria \\
	\texttt{stefanie.elgeti@tuwien.ac.at} \\
}

\renewcommand{\shorttitle}{Automatic CAD Reconstruction of Post-Analysis Geometries}

\hypersetup{
pdftitle={Automatic CAD Reconstruction of Post-Analysis Geometries},
pdfsubject={cs.CE},
pdfauthor={Sebastian ~Hube, Roxana ~Pohlmann, Stefanie ~Elgeti},
pdfkeywords={shape optimization, reverse engineering, CAD, CAE},
}

\DeclareUnicodeCharacter{2212}{-}
\DeclareMathOperator*{\argmin}{arg\,min}
\DeclareMathOperator{\spn}{span}

\newcommand{\mybar}[3]{
	\mathrlap{\hspace{#2}\overline{\scalebox{#1}[1]{\phantom{\ensuremath{#3}}}}}\ensuremath{#3}
}

\newcommand\norm[1]{\left\lVert#1\right\rVert}

\newcommand{\deformed}[1]{\widehat{#1}}
\newcommand{\initialmesh}{\boldsymbol{\mathcal{X}}}
\newcommand{\deformedmesh}{\deformed{\initialmesh}}
\newcommand{\controlpoint}{\mathbf{D}}
\newcommand{\spline}[1]{\boldsymbol{\mathcal{#1}}}
\newcommand{\pointsonspline}{_{\spline{S}}}

\begin{document}
\maketitle

\begin{abstract}
	A key step during industrial design is the passing of design information from computer aided design (CAD) to analysis tools (CAE) and vice versa. 
Here, one is faced with a severe incompatibility in geometry representation: While CAD is usually based on surface representations, analysis mostly relies on volumetric representations.
The forward pass, i.e., converting CAD data to computational meshes, is well understood and established.
However, the same does not hold for the inverse direction, i.e., CAD reconstruction of deformed geometries resulting from analysis. 
This is particularly important for industrial workflows in which the shape optimization of an initial product is outsourced.
Such shape optimization is the focus of this work.
The few reconstruction methods reported mainly rely on spline fitting, in particular on creating new splines similar to shape reconstruction from 3D imaging. 
In contrast, this paper studies a novel approach that reuses the CAD data given in the initial design. 
We show that this concept enables one to shape reconstruct mediocre deformations while preserving the initial notion of features defined during design. 
Furthermore, reusing the initial CAD representation reduces the shape reconstruction problem to a shape modification problem. 
We study this unique feature and show that it enables the reconstruction of CAD data from computational meshes by composing each spline in the initial CAD data with a single, global deformation spline. 
While post-processing is needed for use in current CAD software, most notably, this novel approach enables reconstructing complete CAD models even from defeatured computational meshes.
\end{abstract}


\section{Introduction}
\label{sec_intro}

The current engineering design workflow relies heavily on computer-aided approaches, i.e., CAx. 
CAx is an umbrella term for computer-aided design (CAD) -- i.e., the software utilized for the geometrical design --, 
computer-aided engineering -- i.e., the software utilized for computational analysis --, and computer-aided manufacturing (CAM) 
-- referring to the use of software to guide machine tools. 
Despite the unity suggested by the umbrella term CAx, each individual component has evolved from its unique community, 
leading to significant differences in how data is represented and processed; and this is not only a matter of converting between different formats, 
but instead a matter of completely different concepts of geometry representation, as for example described in \cite{gonzalez2017survey}. 
In practice, this can make the interaction between the different elements of CAx not only tedious, 
but there is also a loss of information along the design pipeline. It is exactly this loss of information that has motivated the method presented in this paper. 
In particular, we will present a method that allows to reconstruct CAD data from a shape-optimization analysis (CAE) under consideration of the initial CAD 
representation.  

To contextualize this work, we start with a summary of the industrial design process, which almost exclusively revolves around CAD. 
The first step of the design process is the system design, from which the conceptual principals are derived. 
These then lead to the requirements on individual components, which can then either be selected from a catalogue or designed individually. 
It is this last step -- the individual component design -- that is the focus of this paper. 
Component design can be based on either design-altering methods, such as shape optimization, 
or design-generating methods, such as topology optimization \cite{allaire2021shape,bletzinger_shape_2017,bendsoe_topology_2003}. 

\begin{figure}[!ht]
	\centering
		\begin{subfigure}[b]{.35\linewidth}
			\tikzset{every picture/.style={line width=0.75pt}} 

\begin{tikzpicture}[x=0.75pt,y=0.75pt,yscale=-0.55,xscale=0.55]

\draw  [fill={rgb, 255:red, 255; green, 255; blue, 255 }  ,fill opacity=1 ] (551.12,80.22) -- (557.52,88.42) -- (544.73,88.42) -- cycle ;
\draw    (541.07,89.63) -- (563,89.63) ;
\draw    (542.9,89.63) -- (541.07,91) ;
\draw    (544.73,89.63) -- (542.9,91) ;
\draw    (546.55,89.63) -- (544.73,91) ;
\draw    (548.38,89.63) -- (546.55,91) ;
\draw    (550.21,89.63) -- (548.38,91) ;
\draw    (552.04,89.63) -- (550.21,91) ;
\draw    (553.86,89.63) -- (552.04,91) ;
\draw    (555.69,89.63) -- (553.86,91) ;
\draw    (557.52,89.63) -- (555.69,91) ;
\draw    (559.35,89.63) -- (557.52,91) ;
\draw    (561.17,89.63) -- (559.35,91) ;

\draw  [fill={rgb, 255:red, 255; green, 255; blue, 255 }  ,fill opacity=1 ] (549.3,80.22) .. controls (549.3,79.46) and (550.11,78.85) .. (551.12,78.85) .. controls (552.13,78.85) and (552.95,79.46) .. (552.95,80.22) .. controls (552.95,80.97) and (552.13,81.58) .. (551.12,81.58) .. controls (550.11,81.58) and (549.3,80.97) .. (549.3,80.22) -- cycle ;

\draw  [fill={rgb, 255:red, 255; green, 255; blue, 255 }  ,fill opacity=1 ] (284.05,80.32) -- (290.45,89.14) -- (277.65,89.14) -- cycle ;
\draw  [fill={rgb, 255:red, 255; green, 255; blue, 255 }  ,fill opacity=1 ] (282.22,80.32) .. controls (282.22,79.51) and (283.04,78.85) .. (284.05,78.85) .. controls (285.06,78.85) and (285.88,79.51) .. (285.88,80.32) .. controls (285.88,81.13) and (285.06,81.79) .. (284.05,81.79) .. controls (283.04,81.79) and (282.22,81.13) .. (282.22,80.32) -- cycle ;
\draw    (274,89.14) -- (295.93,89.14) ;
\draw    (275.83,89.14) -- (274,90.61) ;
\draw    (277.65,89.14) -- (275.83,90.61) ;
\draw    (279.48,89.14) -- (277.65,90.61) ;
\draw    (281.31,89.14) -- (279.48,90.61) ;
\draw    (283.14,89.14) -- (281.31,90.61) ;
\draw    (284.96,89.14) -- (283.14,90.61) ;
\draw    (286.79,89.14) -- (284.96,90.61) ;
\draw    (288.62,89.14) -- (286.79,90.61) ;
\draw    (290.45,89.14) -- (288.62,90.61) ;
\draw    (292.27,89.14) -- (290.45,90.61) ;
\draw    (294.1,89.14) -- (292.27,90.61) ;

\draw  [color={rgb, 255:red, 156; green, 59; blue, 100 }  ,draw opacity=1 ][fill={rgb, 255:red, 156; green, 59; blue, 100 }  ,fill opacity=1 ] (504.56,35.73) -- (550.28,78.85) -- (503.28,78.85) -- (282.77,78.85) -- (330.61,35.73) -- (440.43,35.73) -- cycle ;
\draw    (416.53,6) -- (416.53,28.78) -- (416.53,33) ;
\draw [shift={(416.53,36)}, rotate = 270] [fill={rgb, 255:red, 0; green, 0; blue, 0 }  ][line width=0.08]  [draw opacity=0] (8.93,-4.29) -- (0,0) -- (8.93,4.29) -- cycle    ;
\draw  [color={rgb, 255:red, 255; green, 255; blue, 255 }  ,draw opacity=1 ][fill={rgb, 255:red, 255; green, 255; blue, 255 }  ,fill opacity=1 ] (399.14,57.34) .. controls (399.14,48.94) and (406.92,42.13) .. (416.53,42.13) .. controls (426.13,42.13) and (433.91,48.94) .. (433.91,57.34) .. controls (433.91,65.74) and (426.13,72.56) .. (416.53,72.56) .. controls (406.92,72.56) and (399.14,65.74) .. (399.14,57.34) -- cycle ;
\draw  [color={rgb, 255:red, 255; green, 255; blue, 255 }  ,draw opacity=1 ][fill={rgb, 255:red, 255; green, 255; blue, 255 }  ,fill opacity=1 ] (454.5,57.24) .. controls (454.5,48.84) and (462.28,42.03) .. (471.89,42.03) .. controls (481.49,42.03) and (489.27,48.84) .. (489.27,57.24) .. controls (489.27,65.64) and (481.49,72.46) .. (471.89,72.46) .. controls (462.28,72.46) and (454.5,65.64) .. (454.5,57.24) -- cycle ;
\draw  [color={rgb, 255:red, 255; green, 255; blue, 255 }  ,draw opacity=1 ][fill={rgb, 255:red, 255; green, 255; blue, 255 }  ,fill opacity=1 ] (343.78,57.29) .. controls (343.78,48.89) and (351.56,42.08) .. (361.17,42.08) .. controls (370.77,42.08) and (378.55,48.89) .. (378.55,57.29) .. controls (378.55,65.69) and (370.77,72.51) .. (361.17,72.51) .. controls (351.56,72.51) and (343.78,65.69) .. (343.78,57.29) -- cycle ;

\draw (423,8) node [anchor=north west][inner sep=0.75pt]   [align=left] {{\small F}};

\end{tikzpicture}
		\end{subfigure}
		\hspace{1cm}
		\begin{subfigure}[b]{.35\linewidth}
			\tikzset{every picture/.style={line width=0.75pt}}

\begin{tikzpicture}[x=0.75pt,y=0.75pt,yscale=-0.55,xscale=0.55]

\draw  [fill={rgb, 255:red, 255; green, 255; blue, 255 }  ,fill opacity=1 ] (543.12,80.22) -- (549.52,88.42) -- (536.73,88.42) -- cycle ;
\draw    (533.07,89.63) -- (555,89.63) ;
\draw    (534.9,89.63) -- (533.07,91) ;
\draw    (536.73,89.63) -- (534.9,91) ;
\draw    (538.55,89.63) -- (536.73,91) ;
\draw    (540.38,89.63) -- (538.55,91) ;
\draw    (542.21,89.63) -- (540.38,91) ;
\draw    (544.04,89.63) -- (542.21,91) ;
\draw    (545.86,89.63) -- (544.04,91) ;
\draw    (547.69,89.63) -- (545.86,91) ;
\draw    (549.52,89.63) -- (547.69,91) ;
\draw    (551.35,89.63) -- (549.52,91) ;
\draw    (553.17,89.63) -- (551.35,91) ;

\draw  [fill={rgb, 255:red, 255; green, 255; blue, 255 }  ,fill opacity=1 ] (541.3,80.22) .. controls (541.3,79.46) and (542.11,78.85) .. (543.12,78.85) .. controls (544.13,78.85) and (544.95,79.46) .. (544.95,80.22) .. controls (544.95,80.97) and (544.13,81.58) .. (543.12,81.58) .. controls (542.11,81.58) and (541.3,80.97) .. (541.3,80.22) -- cycle ;

\draw  [fill={rgb, 255:red, 255; green, 255; blue, 255 }  ,fill opacity=1 ] (276.05,80.32) -- (282.45,89.14) -- (269.65,89.14) -- cycle ;
\draw  [fill={rgb, 255:red, 255; green, 255; blue, 255 }  ,fill opacity=1 ] (274.22,80.32) .. controls (274.22,79.51) and (275.04,78.85) .. (276.05,78.85) .. controls (277.06,78.85) and (277.88,79.51) .. (277.88,80.32) .. controls (277.88,81.13) and (277.06,81.79) .. (276.05,81.79) .. controls (275.04,81.79) and (274.22,81.13) .. (274.22,80.32) -- cycle ;
\draw    (266,89.14) -- (287.93,89.14) ;
\draw    (267.83,89.14) -- (266,90.61) ;
\draw    (269.65,89.14) -- (267.83,90.61) ;
\draw    (271.48,89.14) -- (269.65,90.61) ;
\draw    (273.31,89.14) -- (271.48,90.61) ;
\draw    (275.14,89.14) -- (273.31,90.61) ;
\draw    (276.96,89.14) -- (275.14,90.61) ;
\draw    (278.79,89.14) -- (276.96,90.61) ;
\draw    (280.62,89.14) -- (278.79,90.61) ;
\draw    (282.45,89.14) -- (280.62,90.61) ;
\draw    (284.27,89.14) -- (282.45,90.61) ;
\draw    (286.1,89.14) -- (284.27,90.61) ;

\draw  [color={rgb, 255:red, 156; green, 59; blue, 100 }  ,draw opacity=1 ][fill={rgb, 255:red, 156; green, 59; blue, 100 }  ,fill opacity=1 ] (495.72,35.73) -- (542.28,78.85) -- (495.28,78.85) -- (274.77,78.85) -- (322.61,35.73) -- (432.43,35.73) -- cycle ;
\draw  [color={rgb, 255:red, 255; green, 255; blue, 255 }  ,draw opacity=1 ][fill={rgb, 255:red, 255; green, 255; blue, 255 }  ,fill opacity=1 ] (326.99,47.49) .. controls (335.76,43.57) and (351.37,41.12) .. (357.69,43.57) .. controls (364.01,46.02) and (377.61,51.51) .. (362.08,63.17) .. controls (346.55,74.83) and (333.75,71.5) .. (313.84,71.01) .. controls (293.92,70.52) and (318.22,51.41) .. (326.99,47.49) -- cycle ;
\draw  [color={rgb, 255:red, 255; green, 255; blue, 255 }  ,draw opacity=1 ][fill={rgb, 255:red, 255; green, 255; blue, 255 }  ,fill opacity=1 ] (428.26,43.96) .. controls (441.98,46.65) and (438.66,48.73) .. (432.65,59.64) .. controls (426.63,70.56) and (424.19,71.45) .. (410.72,71.4) .. controls (397.25,71.35) and (389,67.3) .. (386,60) .. controls (383,52.7) and (376.2,46.26) .. (388.79,43.96) .. controls (401.39,41.67) and (414.54,41.28) .. (428.26,43.96) -- cycle ;
\draw  [color={rgb, 255:red, 255; green, 255; blue, 255 }  ,draw opacity=1 ][fill={rgb, 255:red, 255; green, 255; blue, 255 }  ,fill opacity=1 ] (480,42.56) .. controls (485.88,45.4) and (504.3,53.24) .. (510.7,62.16) .. controls (517.09,71.08) and (497.29,69.71) .. (480,70) .. controls (462.71,70.29) and (455.62,70.49) .. (453.69,62.16) .. controls (451.75,53.83) and (449.12,45.99) .. (458.07,42.56) .. controls (467.03,39.13) and (474.12,39.72) .. (480,42.56) -- cycle ;
\draw    (408.53,6) -- (408.53,28.78) -- (408.53,33) ;
\draw [shift={(408.53,36)}, rotate = 270] [fill={rgb, 255:red, 0; green, 0; blue, 0 }  ][line width=0.08]  [draw opacity=0] (8.93,-4.29) -- (0,0) -- (8.93,4.29) -- cycle    ;

\draw (415,8) node [anchor=north west][inner sep=0.75pt]   [align=left] {{\small F}};

\end{tikzpicture}
		\end{subfigure}
		\caption{Shape optimization of the initial design (left) modifies the boundaries of existing geometric features (right). Modified from \cite{lang_simultaneous_2021}.}
	\label{fig:shapeOptimization}
\end{figure}
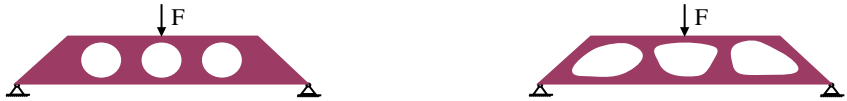

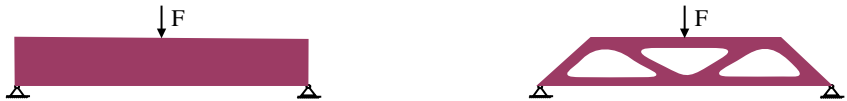
\begin{figure}[!ht]
	\centering
		\begin{subfigure}[b]{.35\linewidth}
			\tikzset{every picture/.style={line width=0.75pt}} 

\begin{tikzpicture}[x=0.75pt,y=0.75pt,yscale=-0.55,xscale=0.55]

\draw  [fill={rgb, 255:red, 255; green, 255; blue, 255 }  ,fill opacity=1 ] (553.12,83.22) -- (559.52,91.42) -- (546.73,91.42) -- cycle ;
\draw    (543.07,92.63) -- (565,92.63) ;
\draw    (544.9,92.63) -- (543.07,94) ;
\draw    (546.73,92.63) -- (544.9,94) ;
\draw    (548.55,92.63) -- (546.73,94) ;
\draw    (550.38,92.63) -- (548.55,94) ;
\draw    (552.21,92.63) -- (550.38,94) ;
\draw    (554.04,92.63) -- (552.21,94) ;
\draw    (555.86,92.63) -- (554.04,94) ;
\draw    (557.69,92.63) -- (555.86,94) ;
\draw    (559.52,92.63) -- (557.69,94) ;
\draw    (561.35,92.63) -- (559.52,94) ;
\draw    (563.17,92.63) -- (561.35,94) ;

\draw  [fill={rgb, 255:red, 255; green, 255; blue, 255 }  ,fill opacity=1 ] (551.3,83.22) .. controls (551.3,82.46) and (552.11,81.85) .. (553.12,81.85) .. controls (554.13,81.85) and (554.95,82.46) .. (554.95,83.22) .. controls (554.95,83.97) and (554.13,84.58) .. (553.12,84.58) .. controls (552.11,84.58) and (551.3,83.97) .. (551.3,83.22) -- cycle ;

\draw  [fill={rgb, 255:red, 255; green, 255; blue, 255 }  ,fill opacity=1 ] (286.05,83.32) -- (292.45,92.14) -- (279.65,92.14) -- cycle ;
\draw  [fill={rgb, 255:red, 255; green, 255; blue, 255 }  ,fill opacity=1 ] (284.22,83.32) .. controls (284.22,82.51) and (285.04,81.85) .. (286.05,81.85) .. controls (287.06,81.85) and (287.88,82.51) .. (287.88,83.32) .. controls (287.88,84.13) and (287.06,84.79) .. (286.05,84.79) .. controls (285.04,84.79) and (284.22,84.13) .. (284.22,83.32) -- cycle ;
\draw    (276,92.14) -- (297.93,92.14) ;
\draw    (277.83,92.14) -- (276,93.61) ;
\draw    (279.65,92.14) -- (277.83,93.61) ;
\draw    (281.48,92.14) -- (279.65,93.61) ;
\draw    (283.31,92.14) -- (281.48,93.61) ;
\draw    (285.14,92.14) -- (283.31,93.61) ;
\draw    (286.96,92.14) -- (285.14,93.61) ;
\draw    (288.79,92.14) -- (286.96,93.61) ;
\draw    (290.62,92.14) -- (288.79,93.61) ;
\draw    (292.45,92.14) -- (290.62,93.61) ;
\draw    (294.27,92.14) -- (292.45,93.61) ;
\draw    (296.1,92.14) -- (294.27,93.61) ;

\draw  [color={rgb, 255:red, 156; green, 59; blue, 100 }  ,draw opacity=1 ][fill={rgb, 255:red, 156; green, 59; blue, 100 }  ,fill opacity=1 ] (552.45,40.48) -- (552.28,81.85) -- (284.77,81.85) -- (284.45,38.48) -- cycle ;
\draw    (418.53,9) -- (418.53,31.78) -- (418.53,36) ;
\draw [shift={(418.53,39)}, rotate = 270] [fill={rgb, 255:red, 0; green, 0; blue, 0 }  ][line width=0.08]  [draw opacity=0] (8.93,-4.29) -- (0,0) -- (8.93,4.29) -- cycle    ;

\draw (425,11) node [anchor=north west][inner sep=0.75pt]   [align=left] {{\small F}};

\end{tikzpicture}
		\end{subfigure}
		\hspace{1cm}
		\begin{subfigure}[b]{.35\linewidth}
			\tikzset{every picture/.style={line width=0.75pt}}

\begin{tikzpicture}[x=0.75pt,y=0.75pt,yscale=-0.55,xscale=0.55]

\draw  [fill={rgb, 255:red, 255; green, 255; blue, 255 }  ,fill opacity=1 ] (556.12,80.22) -- (562.52,88.42) -- (549.73,88.42) -- cycle ;
\draw    (546.07,89.63) -- (568,89.63) ;
\draw    (547.9,89.63) -- (546.07,91) ;
\draw    (549.73,89.63) -- (547.9,91) ;
\draw    (551.55,89.63) -- (549.73,91) ;
\draw    (553.38,89.63) -- (551.55,91) ;
\draw    (555.21,89.63) -- (553.38,91) ;
\draw    (557.04,89.63) -- (555.21,91) ;
\draw    (558.86,89.63) -- (557.04,91) ;
\draw    (560.69,89.63) -- (558.86,91) ;
\draw    (562.52,89.63) -- (560.69,91) ;
\draw    (564.35,89.63) -- (562.52,91) ;
\draw    (566.17,89.63) -- (564.35,91) ;

\draw  [fill={rgb, 255:red, 255; green, 255; blue, 255 }  ,fill opacity=1 ] (554.3,80.22) .. controls (554.3,79.46) and (555.11,78.85) .. (556.12,78.85) .. controls (557.13,78.85) and (557.95,79.46) .. (557.95,80.22) .. controls (557.95,80.97) and (557.13,81.58) .. (556.12,81.58) .. controls (555.11,81.58) and (554.3,80.97) .. (554.3,80.22) -- cycle ;

\draw  [fill={rgb, 255:red, 255; green, 255; blue, 255 }  ,fill opacity=1 ] (289.05,80.32) -- (295.45,89.14) -- (282.65,89.14) -- cycle ;
\draw  [fill={rgb, 255:red, 255; green, 255; blue, 255 }  ,fill opacity=1 ] (287.22,80.32) .. controls (287.22,79.51) and (288.04,78.85) .. (289.05,78.85) .. controls (290.06,78.85) and (290.88,79.51) .. (290.88,80.32) .. controls (290.88,81.13) and (290.06,81.79) .. (289.05,81.79) .. controls (288.04,81.79) and (287.22,81.13) .. (287.22,80.32) -- cycle ;
\draw    (279,89.14) -- (300.93,89.14) ;
\draw    (280.83,89.14) -- (279,90.61) ;
\draw    (282.65,89.14) -- (280.83,90.61) ;
\draw    (284.48,89.14) -- (282.65,90.61) ;
\draw    (286.31,89.14) -- (284.48,90.61) ;
\draw    (288.14,89.14) -- (286.31,90.61) ;
\draw    (289.96,89.14) -- (288.14,90.61) ;
\draw    (291.79,89.14) -- (289.96,90.61) ;
\draw    (293.62,89.14) -- (291.79,90.61) ;
\draw    (295.45,89.14) -- (293.62,90.61) ;
\draw    (297.27,89.14) -- (295.45,90.61) ;
\draw    (299.1,89.14) -- (297.27,90.61) ;

\draw  [color={rgb, 255:red, 156; green, 59; blue, 100 }  ,draw opacity=1 ][fill={rgb, 255:red, 156; green, 59; blue, 100 }  ,fill opacity=1 ] (509.56,35.73) -- (555.28,78.85) -- (508.28,78.85) -- (287.77,78.85) -- (335.61,35.73) -- (445.43,35.73) -- cycle ;
\draw    (421.53,6) -- (421.53,28.78) -- (421.53,33) ;
\draw [shift={(421.53,36)}, rotate = 270] [fill={rgb, 255:red, 0; green, 0; blue, 0 }  ][line width=0.08]  [draw opacity=0] (8.93,-4.29) -- (0,0) -- (8.93,4.29) -- cycle    ;
\draw  [color={rgb, 255:red, 255; green, 255; blue, 255 }  ,draw opacity=1 ][fill={rgb, 255:red, 255; green, 255; blue, 255 }  ,fill opacity=1 ] (516.42,57.15) .. controls (532.42,72.15) and (527.42,70.15) .. (511.42,70.15) .. controls (495.42,70.15) and (482.03,70.38) .. (462.51,70.14) .. controls (442.98,69.91) and (461.42,63.15) .. (475.42,54.15) .. controls (489.42,45.15) and (500.42,42.15) .. (516.42,57.15) -- cycle ;
\draw  [color={rgb, 255:red, 255; green, 255; blue, 255 }  ,draw opacity=1 ][fill={rgb, 255:red, 255; green, 255; blue, 255 }  ,fill opacity=1 ] (403.99,61.33) .. controls (386.98,52.7) and (370.09,45.7) .. (395.57,45.78) .. controls (421.06,45.86) and (421.41,45.7) .. (446.06,45.78) .. controls (470.7,45.86) and (460.11,50.06) .. (441.85,61.33) .. controls (423.58,72.6) and (420.99,69.96) .. (403.99,61.33) -- cycle ;
\draw  [color={rgb, 255:red, 255; green, 255; blue, 255 }  ,draw opacity=1 ][fill={rgb, 255:red, 255; green, 255; blue, 255 }  ,fill opacity=1 ] (371.58,55.21) .. controls (386.74,66.28) and (407.51,70.69) .. (384.2,70.76) .. controls (360.88,70.84) and (344.83,71) .. (325.3,70.76) .. controls (305.77,70.53) and (317.45,66.52) .. (328.45,55.52) .. controls (339.45,44.52) and (356.42,44.15) .. (371.58,55.21) -- cycle ;

\draw (428,8) node [anchor=north west][inner sep=0.75pt]   [align=left] {{\small F}};

\end{tikzpicture}
		\end{subfigure}
	\caption{Topology optimization of the design space (left) determines the material distribution within the design space (right). Modified from \cite{lang_simultaneous_2021}.}
	\label{fig:topologyOptimization}
\end{figure}

\begin{figure}[!ht]
	\centering
		\input{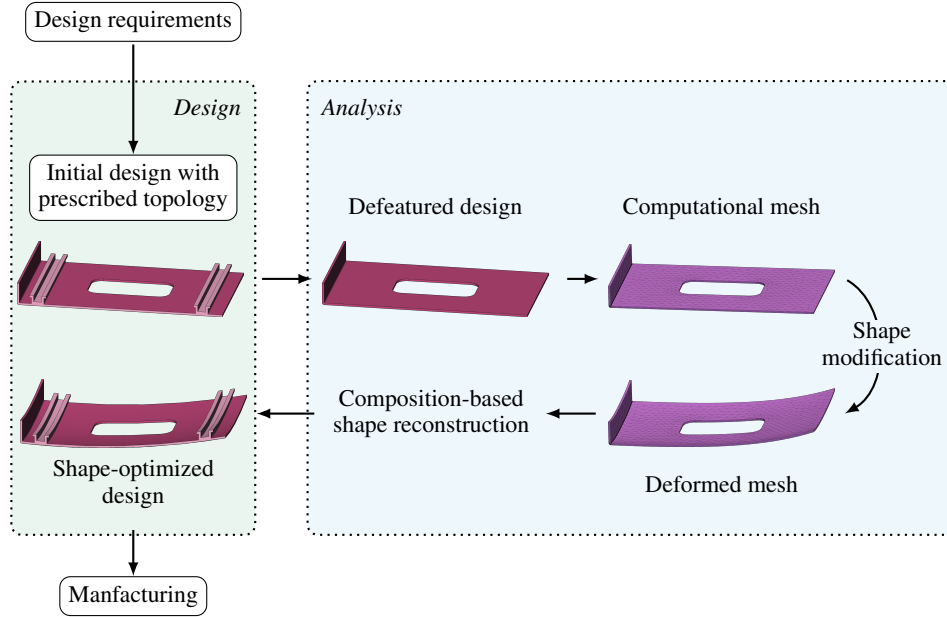}
	\caption{Shape optimization in an industrial design workflow from top to bottom: 
		For efficient computations, the CAD model is defeatured, and a computational 
		mesh is obtained by tesselating this defeatured model. Shape-optimizing CAE yields 
		a deformed computational mesh and this deformation needs to be imposed onto the CAD model.
	}
	\label{fig:problemStatement}
\end{figure}

During shape optimization, the designer creates an experience-driven initial design, 
which is then modified automatically based on computational analysis. 
This means that the basic layout of the component is already known. 
During the shape optimization process, the component is then adapted to optimally fulfill certain design criteria. 
In particular, there will always exist a continuous mapping between the initial and the optimized geometry. 
Instead, during topology optimization, the designer only defines the boundary conditions and an admissible design space. 
Based on the computational analysis, the optimal load path is identified and as such the necessary material distribution. 
In a sense, the component is generated from scratch. Figs. \ref{fig:shapeOptimization} and \ref{fig:topologyOptimization} 
illustrate how during shape optimization, the principal layout of the component remains intact, 
whereas it is the aim of topology optimization to determine this layout. Both approaches have been studied intensively in literature. 
In the field of shape optimization, most existing literature focuses on how the initial shape can be parameterized in 
such a way that the computational-analysis-based adaptation is flexible and able to represent a broad range of shapes while 
the outcome of the optimization is still manufacturable. In the past, two main approaches have evolved: (a) vertex-based representations 
\cite{hojjat2014vertex,le2011gradient} and (b) spline-based representations \cite{zwar2023shape,rozza2013free}. 
Vertex-based representations feature a higher flexibility in the representable shapes, 
but are faced with a need of CAD-reconstruction of the optimized shape. 
In contrast, spline-based approaches can be transferred back into a CAD-system immediately, but they offer less flexibility. 
Notice also that the splines used for optimization almost never coincide with the splines available via CAD and instead require manual reworking. 
In the field of topology optimization, key areas of research are on the one hand geometry representation for generative design, 
such as SIMP or ESO \cite{deaton2014survey}, but on the other hand also the reconstruction of the generated geometries \cite{subedi_review_2020, carraturo_additive_2021}. 
This paper is only concerned with shape optimization and in particular vertex-based shape optimization. 
What is addressed specifically is the topic of CAD-reconstruction. 
In the following, we will describe the workflow of vertex-based shape optimization, addressing challenges and research gaps. 
Figure~\ref{fig:problemStatement} will guide this description. 

As previously mentioned, based on component-design requirements that have been extracted from the system design, 
the designer will prepare an initial design using CAD. Also as mentioned above, during this first step, 
the principal layout (topology) of the component is fully defined and will not be modified during future steps. 
What is also an essential constituent of this first step is that all features of the future component are already defined. 
These include fillets, drill holes, etc.. This full design proposal is then passed on to the analysis software (CAE). 
Due to the lack of consistency between CAD and CAE, two essential modifications need to be made during this step: 
(1) defeaturing, and (2) change of geometry representation. 
Here, defeaturing refers to the fact that all geometrical features that are not immediately relevant to the to-be-conducted computational analysis 
will be removed \cite{foucault2008adaptation}. This becomes necessary as the representation of all features would result in a level of detail that exceeds 
the available computational resources. 
The step of change of geometry representation becomes necessary as CAD and CAE operate on completely different geometrical concepts. 
While CAD is generally based on a boundary representation (B-rep) of the geometry with Non-Uniform Rational B-Splines (NURBS) \cite{rogers2001introduction}, 
CAE most often requires a tessellated volume representation, such as a computational mesh. 
For the sake of completeness, we mention exceptions to using such tesselated, 
volumetric computational meshes such as isogeometric analysis \cite{IGA_Book}, 
which performs computational analysis on analysis-suitable splines, boundary element methods \cite{Beer14} that do not require a volume mesh, and mesh-free methods \cite{Valizadeh2015} that require no mesh at all. 
However, to this day, none of these methods have been able to replace the established representation of a volumetric computational mesh. 
As such, the need of conversion from the full model as B-rep NURBS to a defeatured computational mesh (and later vice-versa) is still omnipresent. 
The forward step, i.e., generating a defeatured geometry and creating a computational mesh is -- possibly tedious -- yet well-understood. 
This computational mesh can then be used to perform the shape optimization. 
Also this step, which is completely enclosed within the analysis framework, is well-understood. 
The research gap lies in the subsequent step: the reconstruction of a CAD geometry from the shape-optimized, i.e., deformed, computational mesh. 
What we propose in order to close this gap is to adapt the initial CAD geometry based on a geometry transformation guided by the deformed computational mesh and performed using functional composition of splines. 
In addition to the pure spline reconstruction, this will allow to automatically reinstate all removed features. 
Before leading over to the main part of the paper, where our reconstruction approach is explained in detail, we will summarize the state of the art for this particular reconstruction problem.

The answers that literature has found so far to the CAD reconstruction problem is almost exclusively based on spline fitting. 
For example, inspired by creating CAD data from 3D imaging with unknown initial shape, 
many works construct generic initial parameterizations that are subsequently adjusted \cite{WEISS200219, ma_nurbs_1998}. 
This approach can be further enhanced by adjusting the generic spline based on properties of the computational mesh; 
e.g. the mesh resolution can be an indicator of high curvature of the initial geometry. 
In such areas, also the spline resolution can be increased \cite{Becker.2011}. 
A reconstruction whose exact representation is closer to the initial design can be obtained by retrieving the original spline surfaces and reconstructing each one individually using an interpolating plane \cite{Louhichi.2009} or regular lattices \cite{Louhichi.2015}.

This work is structured as follows: 
We first recall the most important aspects of vendor-neutral CAD data formats and Non-Uniform Rational B-Splines (NURBS) as well as the principal concept of 
(1) spline fitting and (2) functional composition in Sec.~\ref{sec_splines}. 
In Sec.~\ref{sec_composition}, we introduce the application of a global reconstruction based on spline composition, and we give a demonstration of our method in Sec.~\ref{sec:examples}. 
Finally, in Sec.~\ref{sec_outlook}, we discuss how the initial CAD representation can be modified via one global deformation operator and as such how the entire CAD model can be reconstructed from a defeatured computational mesh.
\section{Basic concepts: CAD representations, NURBS, and related algorithms}
\label{sec_splines}

In this section, we will introduce a number of concepts related to CAD: CAD representation, NURBS, spline fitting, and functional composition. 
We will also illustrate how these concepts affect the aim of CAD reconstruction of shape-optimized computational meshes.

\subsection{CAD representations}
\begin{figure}[!htbp]
	\centering
	\begin{subfigure}{0.39\linewidth}
		\centering
		\includegraphics[width=0.9\linewidth]{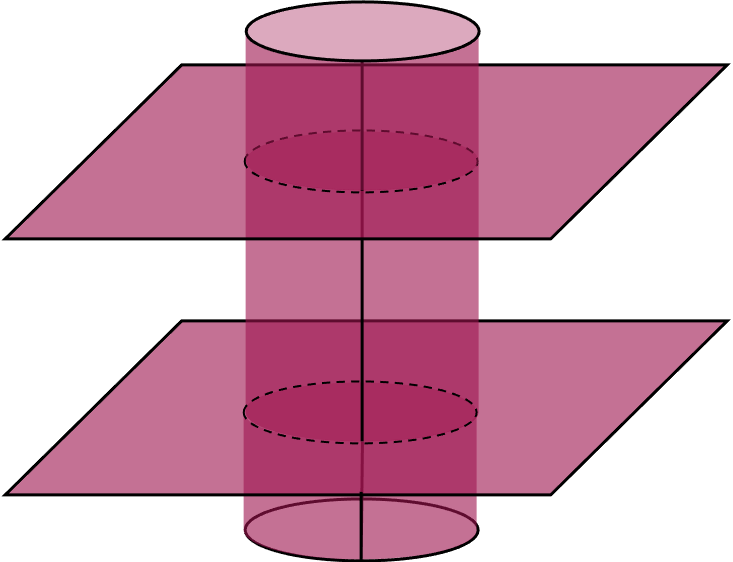}
		\caption{}
		\label{fig:topology_a}
	\end{subfigure}
	\hfill
	\begin{subfigure}{0.29\linewidth}
		\centering
		\includegraphics[width=0.4\linewidth]{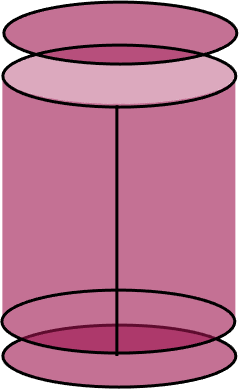}
		\caption{}
		\label{fig:topology_b}
	\end{subfigure}
	\hfill
	\begin{subfigure}{0.29\linewidth}
		\centering
		\includegraphics[width=0.52\linewidth]{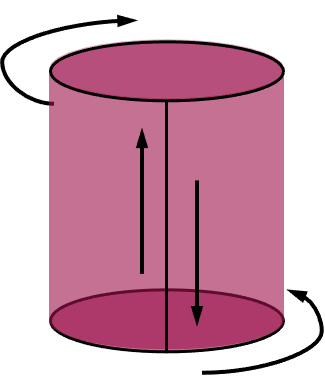}
		\caption{}
		\label{fig:topology_c}
	\end{subfigure}
	\caption{Topological connection of three single surfaces to a closed cylinder in CAD data.
		(A) shows the geometric representation of the single surfaces, (B) the intersected visualized parts, and (C) their topological connection as a surface representation. Modified from \cite{IGES_standard}.}
	\label{fig:topology}
\end{figure}

The internal data structures in a CAD model are comprised of two building blocks: On the one hand, information on individual geometric entities, such as curves or surfaces, is stored, whereas on the other hand, their topological relations (given by Boolean operations) are recorded. 
In the remainder of this work, we refer to these two kinds of data as \textit{geometry data} and \textit{topology data}.
The concept of two kinds of data is visualized in Fig.~\ref{fig:topology}.
\def \subfigwidth {0.3\linewidth}

This particular distinction between geometry and topology allows to update shapes while retaining the topology.
In the example given in Fig.~\ref{fig:topology}, a geometry update could be achieved by offsetting the upper plane.
The resulting cylinder would be higher, but it would still be a cylinder; meaning that the geometry has changed but the topology remained unaffected.
This implies that as long as the topology data remains unaltered and only geometry data is modified, features are unimpaired.\par
Notice that as suggested by Fig.~\ref{fig:topology_b} it is typical that splines comprising a CAD model will extend beyond the actual geometry.
This is related to the fact that CAD design usually mimics manufacturing processes, a fact that cannot be captured if only considering reconstruction based on point clouds.
The novelty of our method is therefore to reuse exactly these initial spline representations as they are native to the CAD software. 
This approach reduces the shape reconstruction problem (cf. Fig.~\ref{fig:problemStatement} bottom) to: \textit{Given post-analysis point data, find the modified CAD geometry data that best approximates these points}.\par
Finally, we would like to remark that not all CAD data is per se represented by splines.
However, we restrict our discussion to vendor-neutral file formats like \textit{IGES}, for which many CAD systems support spline-only export.
Furthermore, converters exist that transform arbitrary geometric representations into splines \cite{NIGES_translator}.

\subsection{A general introduction to NURBS}
\label{subsec:general_spline_intro}
In the next paragraphs, we briefly review the basic concept of splines in general and Non-Uniform Rational B-Splines (NURBS) in particular. 
For further details, the reader is referred to \cite{NURBS_Book}.\par
Since their advent in automotive design, splines became one of the most important geometry parameterizations in CAD.
Splines are interpolants of a set of shape-defining points, referred to as \textit{control points}, $\controlpoint$.
With the basis functions $R_{k,l,m}^{o,p,q}$ of orders $o$, $p$, and $q$ per parametric direction, a volumetric spline body is evaluated as
\begin{equation}
	\spline{T}\left(\xi,\eta,\zeta\right)=\sum_{k=1}^{u}{\sum_{l=1}^{v} \sum_{m=1}^{w} 
	R_{k,l,m}^{o,p,q}\left(\xi, \eta, \zeta\right)\controlpoint_{k,l,m}},
	\label{eq:b_spln_srf}
\end{equation}
where $\xi$, $\eta$, and $\zeta$ denote its parametric coordinates.
A curve $\spline{C}$ or a surface spline $\spline{S}$ is represented analogously by dropping two or one parametric directions.
The basis functions $R_{k,l,m}^{o,p,q}$ are defined on the \textit{parameter space}, which in turn is defined and also partitioned by \textit{knot vectors} $\boldsymbol{\theta}$.
These knot vectors contain a non-decreasing sequence of parametric coordinates, the \textit{knots}.
Given a degree and a knot vector, the basis functions are uniquely defined.
The interval between two knots is referred to as a \textit{knot span}.\par 
Common spline types are B\'ezier splines, B-splines, and NURBS.
Both B\'ezier splines and B-splines feature polynomial basis functions.
What differentiates the two spline types is the support of the basis functions which is always global for the B\'ezier case and can be arbitrarily locally restricted for B-splines.
Instead, NURBS have basis functions that consist of rational polynomials offering an even increased modeling flexibility.

\subsection{Mapping between point clouds using splines}
\label{subsec_SF_intro}

\begin{figure}[hb]
    \centering
	\input{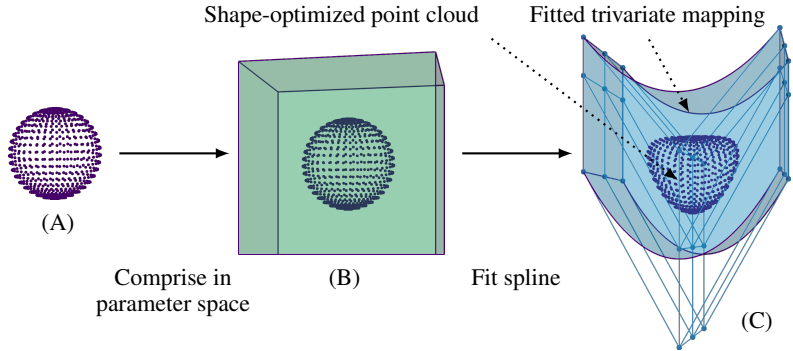}
    \caption{We embed an initial point cloud (A) in a spline's parameter space (B). 
			 By adapting the control points, the trivariate mapping approximates the deformation 
			 between initial and shape-optimized point cloud (C).}
    \label{fig:fittingProcedure}
\end{figure}

In this work, we use splines to map between two point clouds. 
We use spline fitting to construct such a spline mapping from given points, 
where the initial set of points (Fig.~\ref{fig:fittingProcedure}A) is assumed 
to be in the parameter space (Fig.~\ref{fig:fittingProcedure}B). 
We then aim to find a spline representation that interpolates the second set 
of points in physical space (Fig.~\ref{fig:fittingProcedure}C), i.e., after applying the spline mapping.

The starting point of such a spline fit is always a set of sampling points, where for each point the 
coordinates in the initial as well as in the shape-optimized model are known, 
as well as an initial spline.
We associate each target point to one point on the initial spline, i.e., we 
determine the parametric coordinates of the associated point.
Subsequently, we find those control point positions that generate a spline where the distance between the 
target and their associated points is minimized.
Examples of this concept are found in the book by Piegl and Tiller
 \cite{NURBS_Book} or the work by Weiss et al. \cite{WEISS200219}.\par

While the modeling flexibility of a spline largely depends on its parametrization and degree, 
we illustrate the choice of these quantities in Sec.~\ref{sec_composition}, 
and we first outline how to optimally place the control points in the following.
We denote the to-be-deformed spline with a fixed parameterization by $\spline{S}$.
$\initialmesh$ denotes the initial set of grid points that have been generated in such a way that they lie on the initial shape of $\spline{S}$ at the parametric coordinates $\boldsymbol{\Xi}$ meaning that $\spline{S}\left(\boldsymbol{\Xi}\right)=\initialmesh$.
Finally, we are given a set of target points $\deformedmesh$ that are to be interpolated by a deformed version $\deformed{\spline{S}}$.
The Euclidean distance measures the pointwise reconstruction error at point $i$ as
\begin{equation}
R_i = \norm{
	\deformed{
		\spline{S}}\left(
			\boldsymbol{\Xi}_i\right)-\deformedmesh_i} \, .
\end{equation}
In order to determine the new representation, we minimize the summed pointwise reconstruction errors given by
\begin{equation}
	\mathcal{R} = \sum_i R_i = \sum_i{\norm{\deformed{\spline{S}}\left(\boldsymbol{\Xi}_i\right)-\deformedmesh_i}},
	\label{eq:spline_fit_residual}
\end{equation}
where, $\mathcal{R}$ is a direct measure of the reconstruction performance.

We formulate the shape optimization problem: \textit{Find the optimal control points $\controlpoint$ for $\spline{S}$, such that $\mathcal{R}$ is minimized}:
\begin{equation}
	\deformed{\spline{S}_{\controlpoint}} = 
	\underset{
		\controlpoint}{
			\argmin}
	\enskip 
	\sum_i{\norm{\deformed{\spline{S}_{\controlpoint}}\left(\boldsymbol{\Xi}_i\right)-\deformedmesh_i}}
	\label{eq:spline_fit_opt_problem}
\end{equation}
It is beneficial to choose a gradient-based optimization algorithm for this minimization problem since otherwise, undesired results like self-intersecting control polygons would need to be eliminated via constraints on the optimization.
In this work, we solve problem Eq.~\eqref{eq:spline_fit_opt_problem} as an unconstrained optimization problem using the low-storage BFGS algorithm \cite{Nocedal_bfgs1980, Liu_lbfgs1989} implemented in the optimization package NLopt \cite{Johnson2011}.\par
Note that it is also possible to solve the aforementioned spline fitting problem as a linear equation system as:
\begin{equation}
	\deformedmesh = \mathbf{R}\controlpoint,
\end{equation} 
where $\mathbf{R}$ is the matrix of basis functions $R$ (cf. Eq.~\eqref{eq:b_spln_srf}) \textit{evaluated} at the parametric coordinates $\boldsymbol{\Xi}$.

\subsection{Functional composition of splines}
\label{subsec_FC_intro}
\begin{figure}[h]
	\centering
	\begin{subfigure}{.45\linewidth}
		\centering
		\tikzset{every picture/.style={line width=0.75pt}} 
		\begin{tikzpicture}[x=0.75pt,y=0.75pt,yscale=-1,xscale=1]
			\draw (217,105) node  {\includegraphics[width=.7\linewidth]{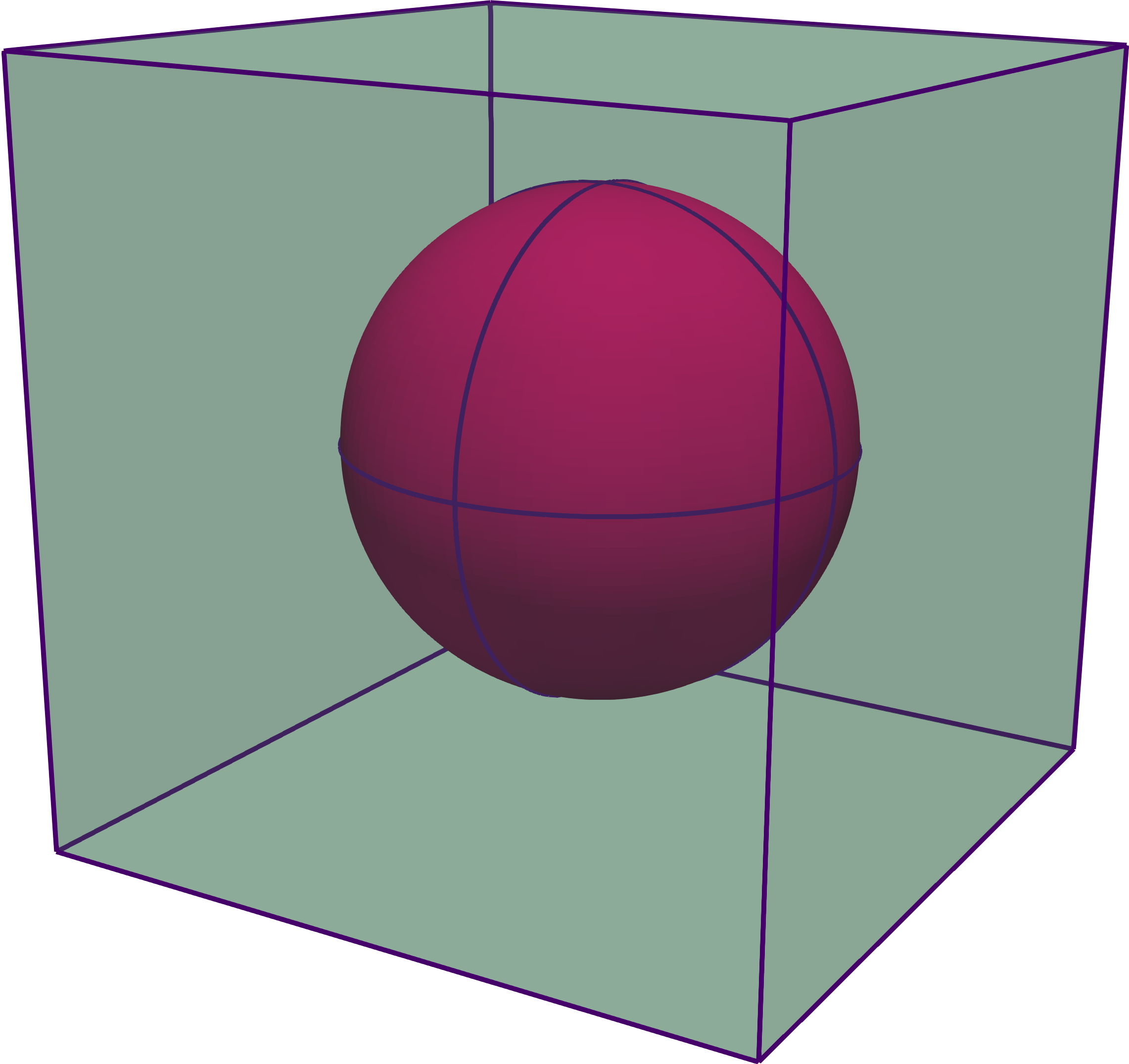}};
			\draw (165,180) node [anchor=north west][inner sep=0.75pt]    {$\xi $};
			\draw (289,165) node [anchor=north west][inner sep=0.75pt]    {$\eta $};
			\draw (110,90) node [anchor=north west][inner sep=0.75pt]    {$\zeta $};	

		\end{tikzpicture}
		\caption{Bi-quadratic sphere $\spline{S}$ placed in parametric domain (green) of $\spline{T}$. }
		\label{fig:sub1}
	\end{subfigure}
	\hfill
	\begin{subfigure}{.45\linewidth}
		\centering
		\includegraphics[width=.7\linewidth]{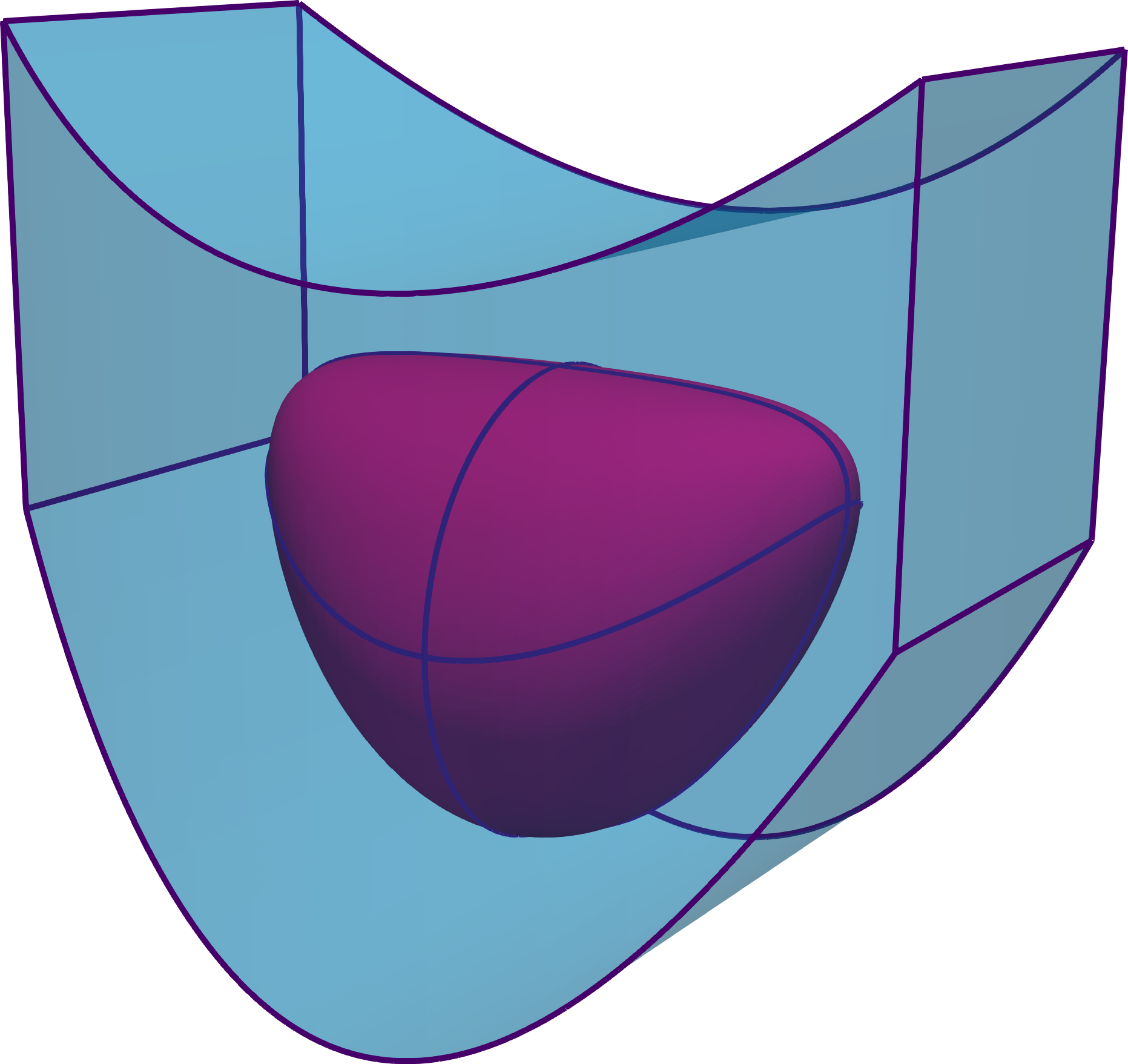}
		\caption{Resulting object $\spline{T} \circ \spline{S}$ shown inside of $\spline{T}$.}
		\label{fig:sub2}
	\end{subfigure}
	\caption{Composition of a sphere and a deformation spline.
		The initial sphere template is bi-quadratic.
		The resulting composed sphere has degrees $o=12$ and $p=12$ along its parametric dimensions.}
	\label{fig:composition}
\end{figure}
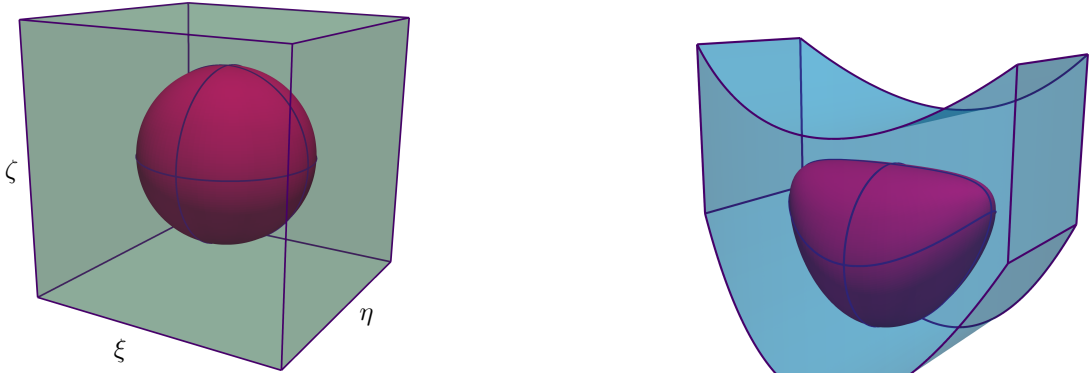

\begin{figure}[h]
	\centering
	\begin{subfigure}[t]{.3\linewidth}
		\centering
		\includegraphics[width=0.9\textwidth]{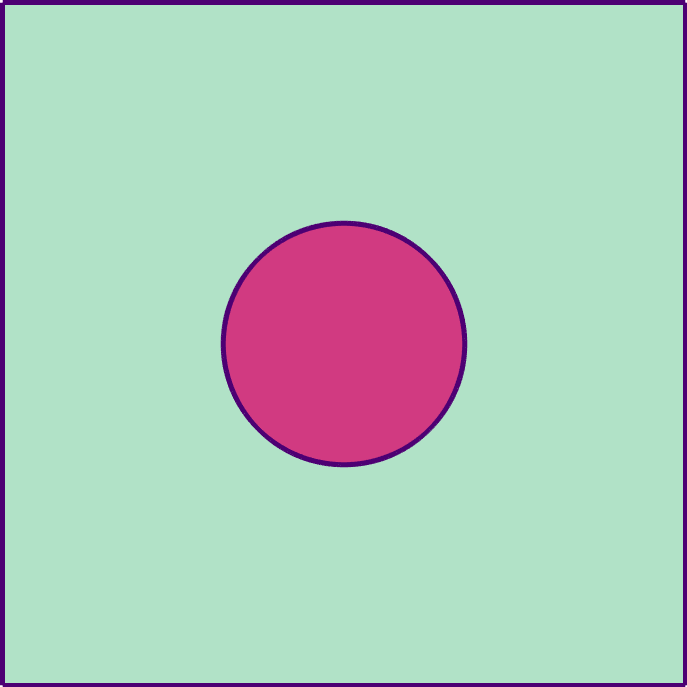}
		\caption{The inner spline (pink) is entirely contained in the only non-zero knot-span (valid).}
		\label{fig:knots1}
	\end{subfigure}
	\hfill
	\begin{subfigure}[t]{.3\textwidth}
		\centering
		\includegraphics[width=0.9\linewidth]{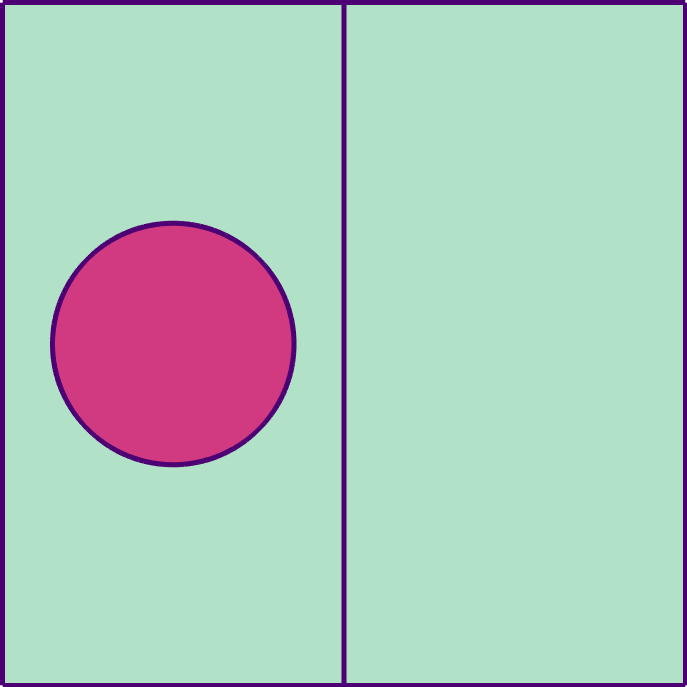}
		\caption{The inner spline (pink) is entirely contained in one of two non-zero knot-spans (valid).}
		\label{fig:knots2}
	\end{subfigure}
	\hfill
	\begin{subfigure}[t]{.3\textwidth}
		\centering
		\includegraphics[width=0.9\linewidth]{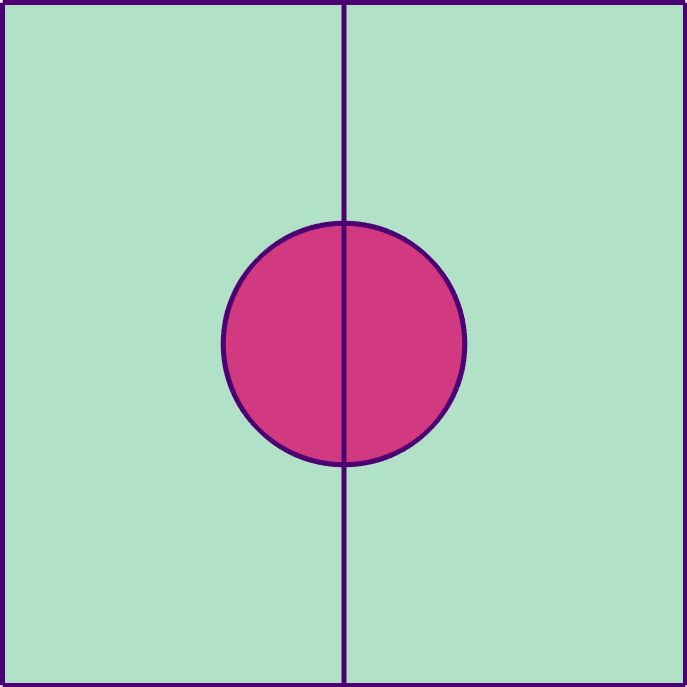}
		\caption{The inner spline (pink) disk crossing a knot line (generally invalid).}
		\label{fig:knots3}
	\end{subfigure}
	\caption{The physical space of a disk (pink) is contained within the parametric space of
		another surface spline (green).}
	\label{fig:knotlines}
\end{figure}

As recently presented in \cite{Elber2016}, two spline mappings can be concatenated as 
\begin{equation}
	\spline{T} \circ \spline{S},
\end{equation}
which can be referred to as \textit{functional composition}.
Effectively, $\spline{S}$ is placed into the parametric domain of $\spline{T}$.
What makes this type of spline manipulation relevant to this paper is that a mapping $\spline{T}$ can be imposed onto $\spline{S}$, leading to a well-defined \textit{deformation} of $\spline{S}$ (cf. Fig.~\ref{fig:composition}).
In general, the resulting geometry is only a valid spline if $\spline{S}$ is entirely contained in one single knot span of $\spline{T}$, meaning that it may not cross knot lines.  
If the outer spline $\spline{T}$ has only one non-zero knot-span (Fig.~\ref{fig:knots1}), this condition is always met. 
In case the outer spline $\spline{T}$ consists of multiple non-zero knot-spans, the inner spline $\spline{S}$ must lie completely within one of those knot spans (Fig.~\ref{fig:knots2}). 
If the inner spline extends beyond one parametric knot-span (Fig.~\ref{fig:knots3}), the composition will -- in general -- not yield a valid result. \par
In the context of this work, we limit ourselves to outer splines with a single non-zero knot-span, which is why techniques 
to overcome this limitation (e.g., \cite{Sosin2018}) have not been pursued.
It may be noted that not using such techniques may, in specific cases, limit the reconstruction accuracy since kinks occurring only in the optimized geometry cannot be exactly captured using a B\'ezier spline.\par

\section{Shape reconstruction using functional composition}
\label{sec_composition}
Existing approaches for CAD/CAE integration that assume an initial CAD model and a deformed mesh \cite{Louhichi.2009, Louhichi.2015}, 
identify and deform each surface entity $\spline{S}_i$ individually. 
In contrast, we approach the problem from a new perspective as one unique volumetric deformation.
This view separates the deformation from the individual surfaces and introduces two tasks: (1) appropriately representing the volumetric deformation, and (2) applying this deformation to the CAD model. 

We suggest to solve these tasks in the following way: First, the deformation encoded in the computational mesh is transferred to one single enclosing spline, a concept similar to free-form deformation \cite{Sederberg1986}.
Subsequently, the deformation is transferred to the initial CAD model by functional composition of the enclosing spline and the splines of the CAD model.
This means that one needs to define the enclosing spline -- a volumetric (trivariate) B\'ezier spline -- that is constructed in such a way that it constitutes a mapping between the initial vertices of the computational mesh $\initialmesh$ and $\deformedmesh$ the new positions of these vertices in the deformed mesh:
\begin{equation}
	\spline{T} : \initialmesh\mapsto \deformedmesh.
	\label{eq:FC_trivar_map}
\end{equation} 
In this case, the sampling points mentioned before are obtained from the initial and deformed computational meshes. This means that we seek for a trivariate that maps the initial computational mesh into the shape-optimized computational mesh. 
This mapping $\spline{T}$ is subsequently applied to \textit{all} entities in the geometry data of the initial CAD model using functional composition such that: 
\begin{equation}
\deformed{\spline{S}}^i = \spline{T} \circ \spline{S}^i\enskip\forall \;i \, .
\end{equation}
Algorithms for functional composition have already been proposed, e.g., by Elber \cite{elber1992} and DeRose et al. \cite{DeRose93}. 
We will thus rather focus on an outline of the approach, and subsequently discuss in detail how we construct the mapping $\spline{T}$.\par
The basic algorithm is given in Algorithm~\ref{alg:FC} and consists of three points: (1) Construct a suitable parametric domain for the trivariate $\spline{T}$, (2) place control points such that $\spline{T}$ fulfills Eq.~\eqref{eq:FC_trivar_map}, and (3) apply functional composition to the initial splines. \par
One particularity of the method is that the functional composition is mathematically exact.
From this exactness follows that the precision of the method is directly governed by how well we fulfill Eq.~\eqref{eq:spline_fit_residual}.\par
	
\begin{algorithm}[!h]
\caption{Shape reconstruction by functional composition}\label{alg:FC}
\begin{flushleft}
	\textbf{Input:} \\
	NURBS-only IGES file;\\
	initial computational mesh $\initialmesh$;\\
	deformed computational mesh $\deformedmesh$;\\
	[1\baselineskip]
	\textbf{Output:} \\
	Updated IGES file;\\
	[1\baselineskip]
	\textbf{Algorithm:} 
\end{flushleft}
\begin{algorithmic}[1]
	\State{$\Sigma = \left\{\spline{S}^i\right\} := \text{Collect all splines from IGES}$};
	\State Construct unitary trivariate $\spline{T}$ such that $\spline{T} \supset \initialmesh$;
	\State 
		\begin{varwidth}[t]{\linewidth}
		$\boldsymbol{\Xi} = \left\{\boldsymbol{\xi}_j \enskip \vert \enskip \spline{T}\left(\boldsymbol{\xi}_j\right) = \mathbf{x}_j \enskip \forall \enskip \mathbf{x}_j \in \initialmesh\right\}$ $:=$ Get parametric coordinates to  $\initialmesh$;
		\end{varwidth}
	\State 
		\begin{varwidth}[t]{\linewidth}
		$\spline{T}_{\deformed{\controlpoint}} = \underset{\controlpoint}{\argmin}\enskip \sum_i{\norm{\spline{T}_{\controlpoint}\left(\boldsymbol{\Xi}_i\right)-\deformedmesh_i }}$  $:=$ Find $\deformed{\controlpoint}$ s.t.  $\spline{T}_{\deformed{\controlpoint}}:\boldsymbol{\Xi}\mapsto\deformedmesh$;
		\end{varwidth}
	\ForAll{$\spline{S}^i \in \Sigma$}
	\State $\deformed{\spline{S}} = \spline{T}_{\deformed{\controlpoint}}\circ\spline{S}$;
	\State Replace $\spline{S}$ by $\deformed{\spline{S}}$ in IGES file;
	\EndFor \\
	\Return Updated IGES file;	
\end{algorithmic}
\end{algorithm}
After this outline of the method, we will now detail on the construction of $\spline{T}$.
To construct $\spline{T}$, we need to define control points $\controlpoint$, knot vectors $\boldsymbol{\theta}$, and degrees per parametric direction.
We start by choosing the degrees.\par
As discussed in Sec.~\ref{subsec_FC_intro}, $\spline{T}$ must be a \textit{B\'ezier spline} (i.e., a single-knot-span B-spline).
Consequently, only the degrees determine the number of control points, and thus influence how well Eq.~\eqref{eq:spline_fit_residual} can be fulfilled.
In other words, $\spline{T}$ needs to be sufficiently fine to capture local shape modifications and since $\spline{T}$ has to be a B\'ezier spline, being \textit{fine} -- in this context -- equals holding a suitably high degree.
In practice, we iteratively solve the optimization problem in Eq.~\eqref{eq:spline_fit_opt_problem} and 
degree-elevate $\spline{T}$ until Eq.~\eqref{eq:spline_fit_residual} is satisfied.\par
With the degrees chosen, 
we require that $\spline{T}$'s parameter space contains all initial points, i.e., $\spn\left(\boldsymbol{\Theta}\right) \supset \initialmesh\pointsonspline$.
Accordingly, we select the knot vectors $\boldsymbol{\Theta}=$$\{\boldsymbol{\theta}_1,$$\boldsymbol{\theta}_2,$$\boldsymbol{\theta}_3\}$ as an equidistantly spaced grid embedded in the linear hull of the initial points.
Choosing the parameter space this way, we obtain the parametric coordinates $\boldsymbol{\Xi}$ (cf. Algorithm~\ref{alg:FC}) from the direct relation
\begin{equation}
	\boldsymbol{\Xi} = \initialmesh.
\end{equation}
For completeness, we shall mention that, in general, we can also utilize Newton's method to determine the parametric coordinates by solving $\spline{T}\left(\boldsymbol{\Xi}_j\right) - \mathbf{x}_j= \boldsymbol{0} \enskip \forall \enskip \mathbf{x}_j \in \initialmesh$.\par
With the parameter space's span determined by $\initialmesh$, we finally place the control points $\controlpoint$ such that $\spline{T}$ prior to the spline fit becomes a unitary mapping.\par

\begin{figure}[ht]
		\centering
		\input{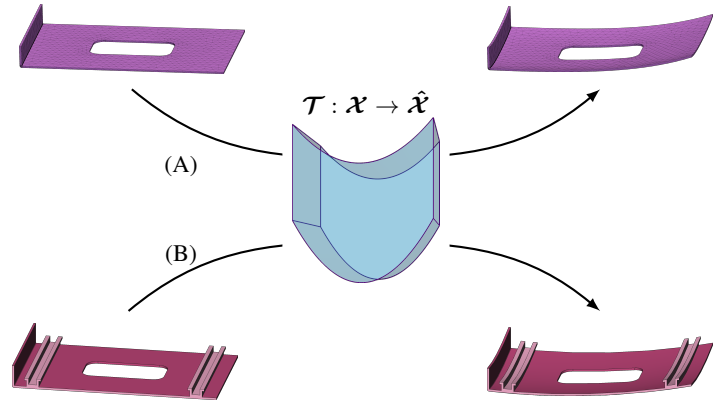}
		\caption{The mapping $\spline{T}$  is defined on the defeatured computational mesh without the stiffeners (A), 
		and applied to the CAD model (B) resulting in updated CAD data.
		The approach yields watertight reconstructions even if defeatured models 
		are used in CAE (compare missing stringers).}
		\label{fig:fitByComposition}
	\end{figure}

{It is essential to note that once fitted, $\spline{T}$ encodes the entire model's deformation as a single continuous mapping. 
From $\spline{T}$'s 's continuity follows that the mapping can be applied to \textit{any} point or spline entirely contained in its parametric space. 
This has the unique effect that one can even reconstruct CAD models from computational meshes, that are topologically non-identical to the original CAD geometry, e.g. caused by defeaturing as shown in Fig.~\ref{fig:fitByComposition}. 	
}

\section{Numerical examples}
\label{sec:examples}
In this section, we apply the proposed shape reconstruction approach based on trivariate fitting 
and spline composition to two numerical examples. 
We start using a plate with a hole as a simplified study example before presenting its application 
to a complex, industrial CAD model.
All calculations were performed serially by an Intel\textsuperscript{\textregistered} Xeon\textsuperscript{\textregistered} 
Platinum 8160  CPU clocked at 2.1\si{\giga\Hz}. 

\subsection{Example: Plate with hole}
	\label{subsec:fc_plate}
	
\begin{figure}[!h]
		\centering
		\includegraphics[width=.6\linewidth]{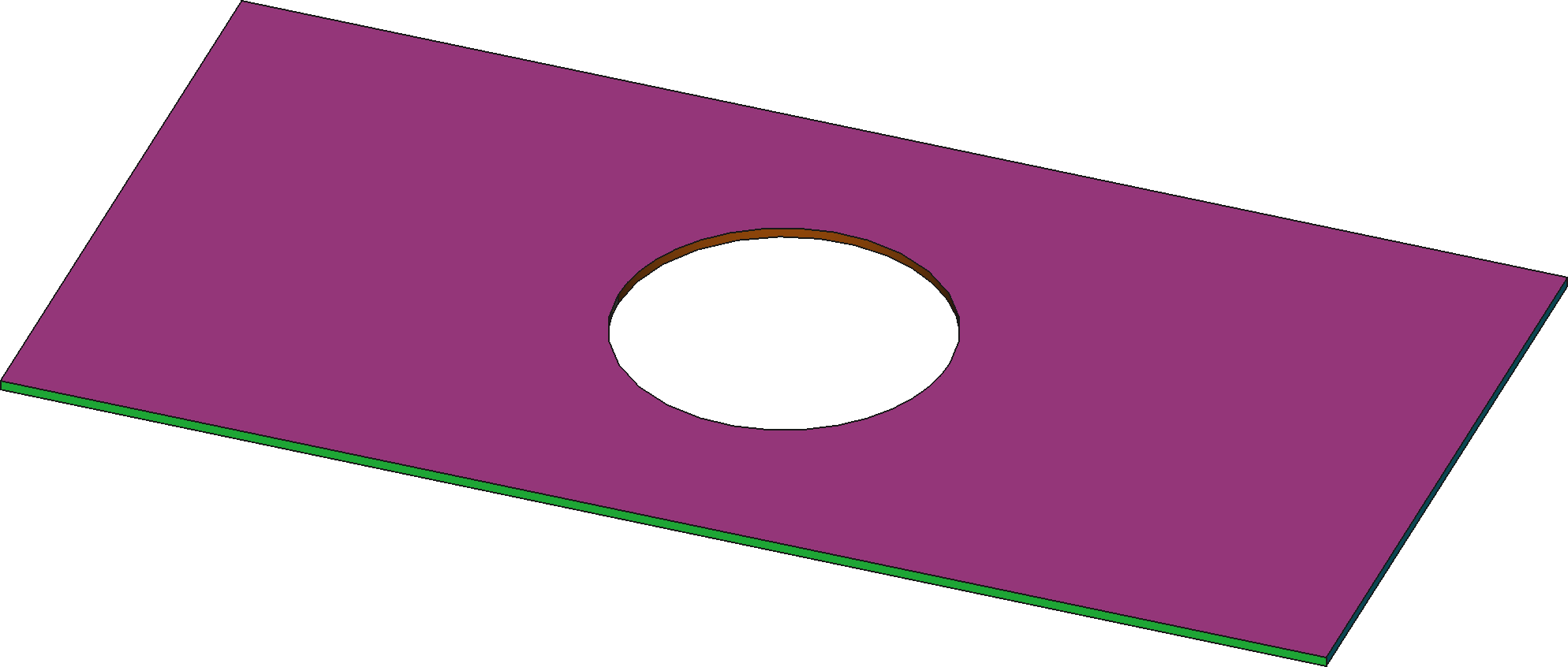}
		\caption{Initial geometry of plate with hole shown with different color for each entity in the geometry data. 
			Results will show the identical coloring for identical entities.
			The dimensions are $\SI{200}{\milli\meter} \times \SI{100}{\milli\meter} \times \SI{1.5}{\milli\meter}$ and the hole diameter is $\num{50}\,\si{\milli\metre}$.}
		\label{fig:res_sf_mesh_plate_initial}
	\end{figure}

	We reconstruct CAD data for a generic plate with dimensions $\SI{200}{\milli\meter} \times \SI{100}{\milli\meter} \times \SI{1.5}{\milli\meter}$ and a central hole with diameter of $50\,\si{\milli\meter}$.
	The geometry is built from six initially planar surfaces, cut by a two-piece cylinder in its center.
	As for the remainder of this work, we plot all entities in the geometry data in individual color, to emphasize that we only change geometry data (cf. Fig.~\ref{fig:topology}).
	The initial CAD representation is shown in Fig.~\ref{fig:res_sf_mesh_plate_initial}.\par
	For the reconstruction, a total of \num{3352} points in $\initialmesh$ is used.
	The prescribed deformation is obtained from a mock simulation result and has no direct physical interpretation.
	The maximal point deflection is $\num{18.88}\,\si{\milli\metre}$.\par
	
\begin{figure}[!h]
		\centering
		\includegraphics[width=.6\linewidth]{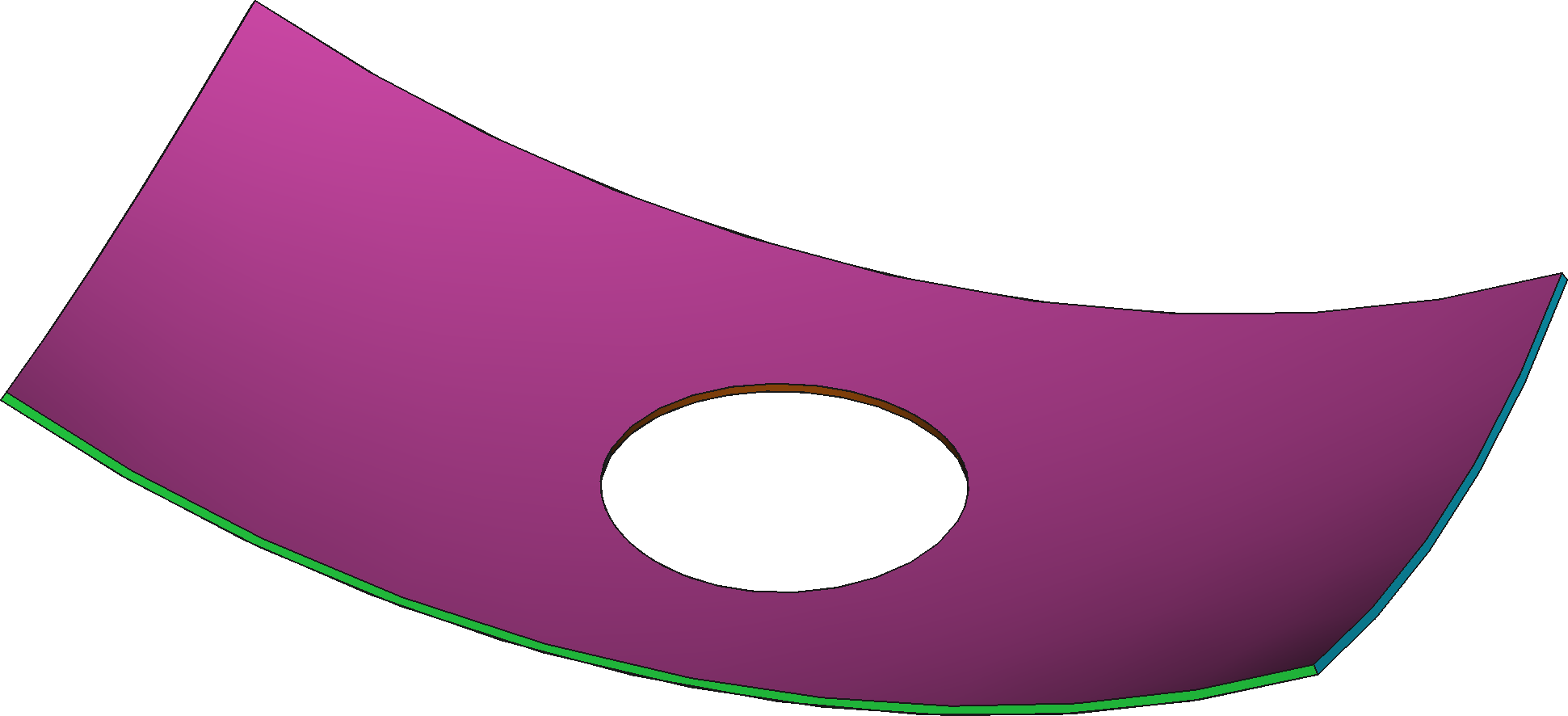}
		\caption{Functional composition applied to a simple geometry.}
		\label{fig:res_fc_nonreduced_plate}
	\end{figure}

	For the reconstruction, we choose degrees $[3,2,1]$ for $\spline{T}$ and obtain the result depicted in Fig.~\ref{fig:res_fc_nonreduced_plate}.
	The resulting spline degrees reach up to 18.
	The reconstruction yields a residual of mean pointwise error 
	$\num{2.96e-1}\,\si{\milli\metre}$, with a maximum point error of $\num{1.08e0}\,\si{\milli\metre}$ (Tab. \ref{tab:plate}). 
	
	As discussed in the construction of $\spline{T}$, increasing the spline degree results in a reduced error.
	For the plate test case with degrees $[19,18,1]$ we obtain a mean error reduction from $\num{2.96e-1}\,\si{\milli\metre}$ to 
	$\num{2.10e-2}\,\si{\milli\metre}$ and maximum error reduction from $\num{1.08e0}\,\si{\milli\metre}$ to $\num{0.15}\,\si{\milli\metre}$ as shown in Tab.~\ref{tab:plate}. 
	The computational time of the reconstruction consists of the time for (a) global splinefit, 
	(b) for the functional composition (FC), and (c) for the low-degree approximation (optional).  
	As Tab.~\ref{tab:plateTime} reveals, fitting $\spline{T}$ dominates the computational effort. 
	The computational time for the functional composition and the low-degree approximation differs largely among individual splines, but generally has a minor effect.
	We compare the computational effort of our global reconstruction method to performing a splinefit for each individual spline in the geometry, related to the approach presented in \cite{Louhichi.2015, Louhichi.2009}.
		
	\begin{table}[h!]
		\centering
		\renewcommand{\arraystretch}{1.3}
		\begin{tabular}{llllll}
			 &   Min. error  & Max. error & Mean error & RMS \\
			\hline			 
			A & $\num{7.5102e-03}$ & $\num{1.0764}$ & $\num{2.9600e-1}$&  $\num{3.5253e-1}$ \\
			B & $\num{7.5103e-03}$ & $\num{1.0764}$ &  $\num{2.9600e-1}$ & $\num{3.5254e-1}$ \\			
			\vspace{0.5pt}
		\end{tabular}	
		\caption{The reconstruction error for the plate with a hole (in \si{\milli\meter}) of (A): the composed model, and (B): the composed and degree-reduced model (cf. Sec~\ref{subsec:loa}).}
		\label{tab:plate}
	\end{table}

	\nprounddigits{3}
	\begin{table}[ht]
		\centering
		\begin{tabular}{llll}
			Process step & Cumulative & Minimal & Maximal \\
			\hline
			Splinefit   	   & $\np{20.0}$ & -- & -- \\
			FC 				   & $\np{0.5641}$ & $\np{6.6640e-3}$ & $\np{6.4384e-2}$\\
			Low-degree approx. & $\np{3.1392}$ & $\np{6.6098e-2}$ & $\np{0.6573}$\\
			\textit{Overall}  & $\np{23.7033}$ & -- & -- \\ 
			\hline 
			Individual splinefit (reference) &$\np{4.303805e4}$ & $\np{20.01}$ & $\np{1.7458e4}$\\
			\vspace{0.5pt}
		\end{tabular}
		\caption{
			Computational time (in \si{\second}) for the plate with a hole. Each major processing step (functional composition, trivariate fit and low-degree approximation) is individually timed for each single spline $\spline{S}^i \in \Sigma$.
			A comparison is given by individually fitting each spline $\spline{S}^i $ instead of using a global deformation spline (Individual splinefit).
		 \label{tab:plateTime}
		}
	\end{table}
	\npnoround

\subsection{Example: Sensor box}
	\label{subsec:fc_sensor}
\begin{figure}[h]
		\centering
		\begin{subfigure}[t]{.45\linewidth}
			\includegraphics[width=\linewidth]{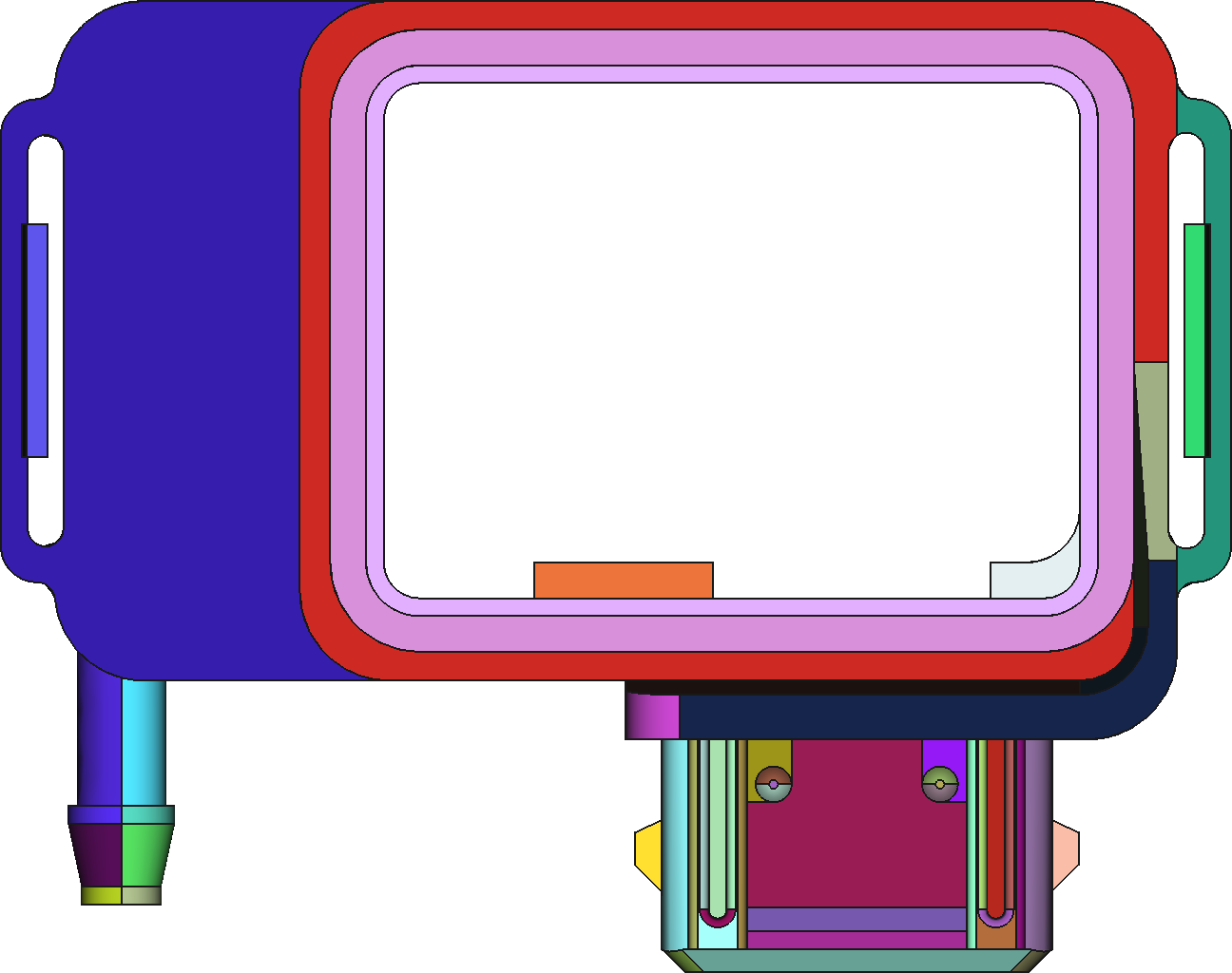}
			\caption{Initial sensor box shown with different color for each entity in the geometry data. Results will show the identical coloring for identical entities.}
			\label{fig:res_sf_mesh_sensor_initial}
		\end{subfigure}
		\hfill
		\begin{subfigure}[t]{.45\linewidth}
			\includegraphics[width=\linewidth]{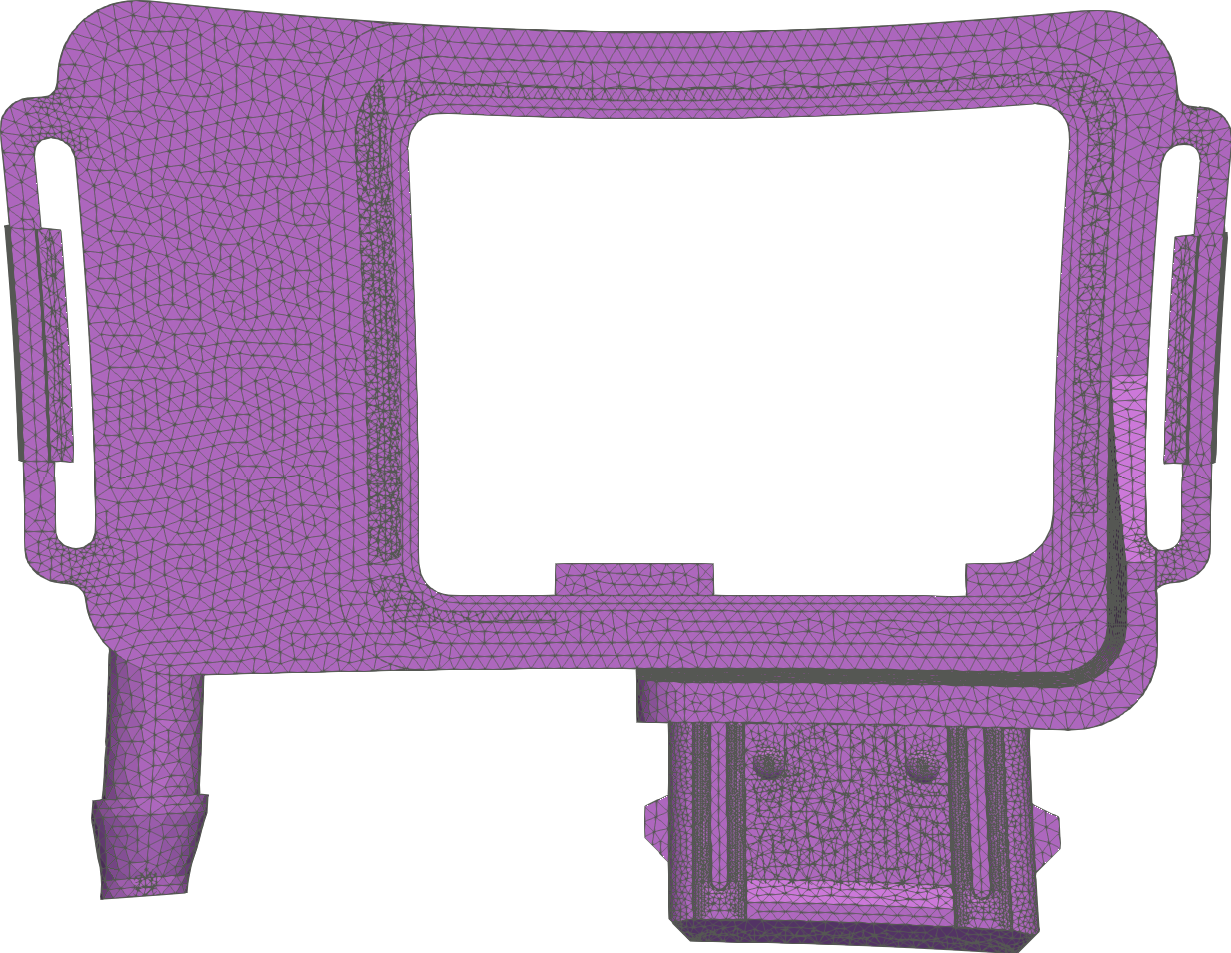}
			\caption{The deformation described in Eq.~(\ref{eq:prescribed_deformation_sensor}) modifies each node's position 
			depending on its distance to the geometric center.}
			\label{fig:sensor_deformed}
		\end{subfigure}
		\caption{The proposed method deforms an initial CAD model (A) to the shape of a deformed computational grid (B).}
	\end{figure}

	As a second and both computationally and methodologically more challenging example, we reconstruct CAD data for a pressure sensor box.
	The initial geometry has length, width and depth of $\mathcal{O}\left(10\right)\,\si{\milli\metre}$ and is shown in Fig.~\ref{fig:res_sf_mesh_sensor_initial}.
	For the reconstruction, a total of $\num{26177}$ grid points is used.
	In absence of simulation data, we prescribe a non-linear deformation (Fig.~\ref{fig:sensor_deformed}) resulting in nodal offsets $\mathbf{d}_j$ 
	for all $\mathbf{x}_j \in \initialmesh$  with $c=0.15$ via:
	\begingroup
	\allowdisplaybreaks[0]

	\begin{subequations}\label{eq:prescribed_deformation_sensor}
		\begin{alignat}{3}
			\mathbf{x}_{\mathrm{center}} &= \mybar{0.6}{1pt}{\initialmesh}\\
			\mathbf{r} &= \left\{r_j \enskip \big\vert \enskip r_j = \left(\mathbf{x}_j-\mathbf{x}_{\mathrm{center}}\right)^2\right\}\\
			\mathbf{d}_j &= c \frac{r_j}{\underset{j}{\max\,}{r_j}}\left(\mathbf{x}_j-\mathbf{x}_{\mathrm{center}}\right),
		\end{alignat}
	\end{subequations}
	\endgroup
	which yields a maximum point deflection of $6.59\,\si{mm}$ and an average deflection of $1.70\,\si{\milli\meter}$.
	
	With degrees $[3, 3, 2]$ for $\spline{T}$, a reconstruction with a mean error of $\num{3.52e-3}\,\si{\milli\meter}$ and a maximum point error 
	of $\num{4.58e-2}\,\si{\milli\meter}$ is achieved as shown in Tab.~\ref{tab:technikum}.
	With initially bicubic splines in the geometry data, the composed splines, however, yield degrees up to 24.
	For the pressure sensor, the mean error reduces from $\num{3.52e-3}\,\si{\milli\meter}$ to $\num{7.11e-4}\,\si{\milli\meter}$, and the maximum error 
	reduces from $\num{4.58e-2}\,\si{\milli\meter}$ to $\num{1.02e-2}\,\si{\milli\meter}$ when a refined spline with degrees $[7,3,4]$ is used.
	Table~\ref{tab:sensorTime} lists the computational time for the individual steps of the reconstruction process.
	
	\begin{table}[h!]
		\centering
		\renewcommand{\arraystretch}{1.3}
		\begin{tabular}{llllll}
			 &   Min. error & Max. error & Mean error & RMS \\
			\hline
			A & $\num{3.3194e-05}$ & $\num{4.5816e-2}$ & $\num{3.5244e-3}$&  $\num{5.2402e-3}$ \\
			B & $\num{3.3213e-5}$ & $\num{4.5816e-2}$ &  $\num{3.5373e-3}$ & $\num{5.2446e-3}$ \\			
			\vspace{0.5pt}
		\end{tabular}	
		\caption{The reconstruction error (in \si{\milli\meter}) for the pressure sensor of: (A) 
		composed model, (B) composed and degree-reduced model (cf. Sec~\ref{subsec:loa}).}
		\label{tab:technikum}	
	\end{table}

	\nprounddigits{3}
	\begin{table}[ht]
		\begin{center}
		\begin{tabular}{llll}
			Process step & Cumulative & Minimal & Maximal \\
			\hline
			Global splinefit   & $\np{5.7853e3}$ & -- & -- \\
			FC 			& $\np{16.1198}$ & $\np{1.0941e-3}$ & $\np{7.8385e-2}$\\
			Low-degree approx. &$\np{4.3431e4}$ & $\np{6.3397}$ & $\np{6.3410e2}$\\
			\textit{Overall}  & $\np{4.9232e4}$ & -- & -- \\ 
			\hline 
			Individual splinefit (reference) &$\np{4.192478e5}$ & $\np{20.01}$ & $\np{1.16611e4}$\\
			\vspace{0.5pt}
		\end{tabular}
		\end{center}
		\caption{
			Computational time (in \si{\second}) for the pressure sensor with our method at each individual step and overall.
			Similar to Tab.~\ref{tab:plateTime}, timings are compared to an individual splinefit of each curve or surface spline in the geometry (Individual splinefit).}
		\label{tab:sensorTime}
	\end{table}
	\npnoround

\subsection{Low-degree approximation}
	\label{subsec:loa}
	One downside of functional composition is the resulting high degree.
	As shown by Elber \cite{Elber2016}, the degrees of a surface $\spline{S}$ of degrees $[m_1, m_2]$ composed with a trivariate $\spline{T}$ of degrees $[n_1, n_2, n_3]$ are
	\begin{equation}
		\left[m_1\left(n_1+n_2+n_3\right), m_2\left(n_1+n_2+n_3\right)\right].
	\end{equation}
	For a fourth degree trivariate composed with a bi-quadratic surface, this already yields degrees $[24,24]$.
	At the same time, we find that most CAD systems are incapable of handling data of such high degrees.\par 
	For such cases, we adopt a low-degree approximation explained in the following.
	Using iterative knot insertion and degree-reduction, we approximate the composed spline by a lower-degree version.
	We build the \textit{degree-reduction} algorithm utilized here around the core algorithm presented in \cite{NURBS_Book}.
	The full algorithm, however, is given in Algorithm~\ref{alg:low-order}.
	Particularly, we measure the error introduced by the degree-reduction and insert additional knots into the knot span that causes the largest error.
	We also observe that the degree-reduction algorithm in \cite{NURBS_Book} tends to produce oscillatory control polygons, which leads to problems during repeated degree-reduction.
	To counteract this behavior, we finally add control points through knot insertion  in every knot span in which the control polygon's length increases more than a selected threshold.
	The additional modeling flexibility thereby introduced, effectively ensures repeated degree-reduction is possible.
	
\begin{figure}[h]
		\centering
		\includegraphics[width=.5\linewidth]{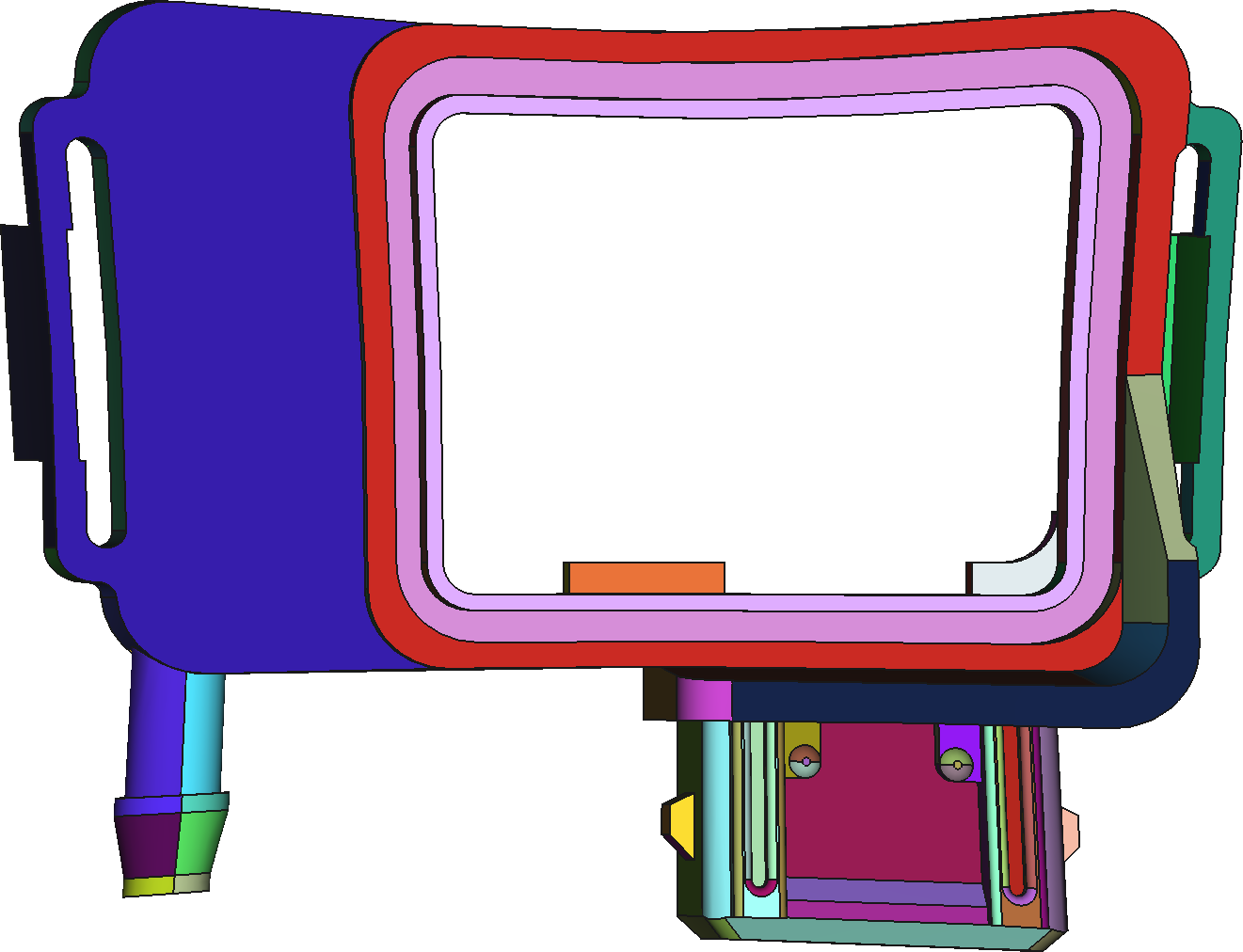}
		\caption{Resulting reconstructed shape with degree-reduction.}
		\label{fig:res_fc_reduced_sensor_overview}
	\end{figure}

	As a demonstration, we apply degree reduction until $[p,q]=[4,4]$ to every curve and surface of both deformed example geometries, where degree four is chosen to provide $G^2$ continuity.		
	For the plate with hole, the reduction increases the mean pointwise error of $\num{2.96e-1}\,\si{\milli\meter}$ by $\num{1.7e-6}\,\si{\milli\meter}$, 
	and the maximum point error of $\num{1.08e0}\,\si{\milli\meter}$ by $\num{0.9e-8}\,\si{\milli\meter}$. \\
	Applying the algorithm to the sensor box, the mean pointwise reconstruction error rises from $\num{3.52e-3}\,\si{\milli\meter}$ to $\num{3.54e-3}\,\si{\milli\meter}$, and the maximum point error of $\num{4.58e-2}\,\si{\milli\meter}$ increases by $\num{1.0e-8}\,\si{\milli\meter}$.
	The reconstruction yields the geometry depicted in Fig.~\ref{fig:res_fc_reduced_sensor_overview}.

\section{Discussion and outlook}
\label{sec_outlook}
This paper introduces a new approach to spline-based CAD reconstruction.
Instead of creating new parameterizations, we recycle spline parameterizations available from initial CAD files.

A composition-based 
approach is presented, which not only appears intuitive as it applies the same 
deformation that has been applied to the mesh during the analysis step also to 
the CAD data, but also presents two unique features: 
(1) Due to the analytically exact shape modification, the initial CAD 
model's level of watertightness is largely preserved, and (2) the deformation of an arbitrary number of 
splines is possible irrespective of the availability of local reconstruction data 
within the mesh. \par
From the industrial view, we shall mention that the composition 
approach is currently limited by two factors:
First, a post-processing, low-degree approximation is needed, which may cause a 
slight loss of watertightness.
One could, however, argue that all trimmed CAD data inherently suffer from 
geometric approximations and inaccuracies \cite{marussig2018}.
This very fundamental shortcoming may invalidate the argument of possible 
inaccuracies resulting from low-degree approximations.
The second factor is its limitation to moderate, smooth deformations as a result of the 
trivariate $\spline{T}$ having to be a B\'ezier spline:
Since fitting very-high-degree B\'ezier splines requires an impractical number of 
control points, the resulting limit on the admissible degree also limits the 
method's ability to recover \textit{local} shape modifications.
\par

However, the significant difference to all state-of-the-art methods is that 
we achieve the reconstruction using a single operator, which we can apply to 
\textit{all} entities in the \textit{initial} CAD data.
From the ability to deform all splines contained in the original CAD file 
follows that composition-based CAD reconstruction is without modification able 
to reconstruct entire CAD models even from post-analysis results obtained 
from defeatured geometries (cf. Fig. \ref{fig:fitByComposition}).
\par
The novel approach of reusing the existing CAD representation, however, not only 
tightens the connection between CAD and CAE but, beyond that, turns out to allow 
the reconstruction of defeatured geometries -- an application outside the practical 
realm of all existing spline fitting methods.
We claim that this unique feature motivates further research towards locally enhanced composition 
methods, e.g., hierarchical application and improved degree-reduction techniques.
\section*{Acknowledgements}
Implementation was done on the HPC cluster provided by IT Center at RWTH Aachen.
Simulations were performed with computing resources granted by RWTH Aachen University under the project thes0991.
All geometry plotting has been done from Pointwise and FreeCad \cite{FreeCAD} software.
Resulting IGES files are accessible under the public project RWTH-2023-05453 \cite{ourdataset}.
\newpage
\section*{Appendix} 
\hphantom{phantom}

\begin{algorithm}[H]
    \begin{flushleft}
	\textbf{Input:}\\
	Spline $\spline{S}$;\\
	Target degree $p$;\\
	\vspace{5mm}
	\textbf{Output:} \\
	Degree-reduced spline $\deformed{\spline{S}}$ with maximal error $\|\deformed{\spline{S}} - \spline{S} \| \leq \varepsilon$;\\
	\vspace{5mm}
	\textbf{Algorithm:} 
    \end{flushleft}
	\begin{algorithmic}[1]
        \caption{Low-degree approximation}\label{alg:low-order}
		\State{Normalize all control point weights to $[0,1]$};
		\While{$\spline{S}.\mathrm{degrees} > p$}
		\State{Calculate length of control polygon};
		\State{
			\begin{varwidth}[t]{\linewidth}
				$\varepsilon, \deformed{\spline{S}} = \spline{S}.\textrm{reduce\_degree()}$ \enskip:= Reduce spline degree with error $\varepsilon$;
			\end{varwidth}
		}
		\While{$\varepsilon > \epsilon$}
		\State{
			\begin{varwidth}[t]{\linewidth}
				Determine the knot span $i$ with the highest error contribution;
		\end{varwidth}}
		\State{
			\begin{varwidth}[t]{\linewidth}
				Insert a new knot $\spline{S}.\textrm{degrees}-p+1$ times at $\frac{1}{2}\left(\xi_{i}+\xi_{i+1}\right)$; 
		\end{varwidth}}
		\State{$\varepsilon, \deformed{\spline{S}} = \spline{S}.\textrm{reduce\_degree()}$}\enskip:= Recompute degree-reduction;
		\EndWhile
		\State{Calculate length of new control polygon};
		
		\If {
				$\frac{\text{length of new control polygon}}{\text{length of control polygon}} > 1.1$
		}
		
		\ForAll{knot spans in $\spline{S}.\textrm{knot spans}$}
		\If {$\frac{\text{length of new \textit{knot span} control polygon}}{\text{length of \textit{knot span} control polygon}} > \delta$}
		\State{Insert a knot in knot span center};
		\EndIf
		\EndFor
		\EndIf
		\State{$\spline{S}=\deformed{\spline{S}}$};
		\EndWhile\\
		\Return $\deformed{\spline{S}}$;
        \end{algorithmic}
\end{algorithm}
\newpage

\bibliographystyle{unsrtnat}  
\bibliography{mybibfile}

\end{document}